\documentclass[fleqn,usenatbib]{mnras}

\usepackage[T1]{fontenc}
\usepackage{ulem}
\DeclareRobustCommand{\VAN}[3]{#2}
\let\VANthebibliography\thebibliography
\def\thebibliography{\DeclareRobustCommand{\VAN}[3]{##3}\VANthebibliography}

\usepackage{graphicx}	
\usepackage{amsmath}	
\usepackage{amssymb}	
\usepackage{newtxtext,newtxmath}
\usepackage{xcolor}
\usepackage{float}

\newcommand{\fhn}[1]{{\textcolor{magenta}{#1}}}

\title[Cosmic web in non-linear regime]{Structure of cosmic web in non-linear regime: the nearest neighbour and spherical contact distributions}

\author[Ansari Fard et al.]{
Mohammad AnsariFard$^{1}$, 
Zahra Baghkhani$^{1}$,
Laya Ghodsi$^{1,2}$,
Sina Taamoli$^{1,3}$,
\newauthor
Farbod Hassani$^{4}$,
Shant Baghram$^{1}$ \thanks{baghram@sharif.edu}
\\ \\
$^{1}$
Department of Physics, Sharif University of
Technology, P.~O.~Box 11155-9161, Tehran, Iran\\
$^{2}$
Department of Physics and Astronomy, University of British Columbia, Vancouver, BC, V6T 1Z1, Canada \\
$^{3}$
Department of Physics and Astronomy, University of California, Riverside, CA 92521, USA \\
$^{4}$
Institute of Theoretical Astrophysics, University of Oslo, 0315 Oslo, Norway \\
}

\date{Accepted XXX. Received YYY; in original form ZZZ}

\pubyear{2022}

\begin{document}
\label{firstpage}
\pagerange{\pageref{firstpage}--\pageref{lastpage}}
\maketitle
\begin{abstract}
In non-linear scales, the matter density distribution is not Gaussian. Consequently, the widely used two-point correlation function is not adequate anymore to capture the matter density field's entire behaviour. {{Among all statistics beyond correlation functions, the spherical contact (or equivalently void function), and nearest neighbour distribution function seem promising tools to probe matter distribution in non-linear regime. In this work, we use halos from cosmological N-body simulations, galaxy groups from the volume-limited galaxy group  and central galaxies from mock galaxy catalogues, to compare the spherical contact with the nearest neighbour distribution functions. We also calculate the J-function (or equivalently the first conditional correlation function), for different samples. Moreover, we consider the redshift evolution and mass-scale dependence of statistics in the simulations and dependence on the magnitude of volume-limited samples in group catalogues as well as the mock central galaxies.}}
The shape of the spherical contact probability distribution function is nearly skew-normal, with skewness and kurtosis being approximately 0.5 and 3, respectively. On the other hand, the nearest neighbour probability distribution function is nearly log-normal, with logarithmic skewness and kurtosis being approximately 0.1 and 2.5, respectively. 
Accordingly, the spherical contact distribution function probes larger scales compared to the nearest neighbour distribution function, which is influenced by details of structures. 
We also find a linear relation between the mean and variance of the spherical contact probability distribution function in simulations and mock galaxies, which could be used as a distinguishing probe of cosmological models.
\end{abstract}

\begin{keywords}
(cosmology:) large-scale structure of the Universe -- (cosmology:) dark matter -- galaxies: statistics
\end{keywords}



\section{Introduction}
The standard model of cosmology is deduced from the early and late time statistics of matter and radiation distribution in the Universe. The two-point correlation function and its wide usage in cosmology are well known. As an example, cosmologists employed the two-point correlation function in the early time Cosmic Microwave Background (CMB) fluctuations \citep{Aghanim:2018eyx,hinshaw2013nine} and the late time Large Scale Structure (LSS) data sets \citep{Sanchez:2012sg, wang2013sdss,shi2016mapping, Alam:2016hwk, ivanov2020cosmological}. The main advantage of the two-point correlation function is in its easy procedure to compare theory with observations. It is used to find the cosmological parameters \citep{Alam:2016hwk,Abbott:2017wau} or even to find a deviation from the standard model of cosmology and to constrain the inflationary models \citep{Akrami:2018odb}. In \cite{Fard:2017oex} some of us studied the specific features (oscillatory and step-function) of the primordial power spectrum and their imprints on early and late-time two-point correlation functions. \\
The study of the LSS in non-linear scales is essential as it is tightly connected to the essence of dark matter. The small scale challenges of cold dark matter could be a hint beyond the standard model of cosmology \citep{deMartino:2020gfi}. The distribution of the dark matter halos and sub-halos in small scales is crucial to test cosmological models. This distribution can be studied by their gravitational effects \cite{Baghram:2011is,Li:2012qha,Rahvar:2013xya,Feldmann:2013hqa} or/and the distribution of satellite galaxies \cite{Agustsson:2015iuo}.\\
As a result of mode coupling in strongly non-linear scales, the matter density distribution is far from Gaussian \citep{Bernardeau:2001qr}. To extract all data's information, we need to investigate statistics, which depend on all higher-order correlations. In this direction, several methods and different quantities have been proposed.
The first choice is to study the higher-order $n$-point correlation functions, directly \citep{scoccimarro1998nonlinear,Verde:2001pf,sefusatti2006cosmology}. However, in strongly non-linear scales, it is compelling to introduce new approaches to extract information from data while reducing computational costs. 
Two approaches are considered to study cosmological structure formation: continuous random field and discrete point processes.

In the context of continuous random field process, the topological structure of the cosmic web has been explored through the genus statistics of iso-density contours \citep{gott1986sponge, hamilton1986topology, melott1990topology} and the Minkowski functionals \citep{Schmalzing:1997aj}. In this direction, the genus statistics is used to constrain the cosmological parameters \citep{iii20083d, Appleby:2018jew, Colley:2014nna}, to test modified gravity theories \citep{wang2012topology}, and dark energy models \citep{zunckel2011using}.

In the second approach, which is the main focus of this work, we consider the matter density as a discrete point process consists of particles, halos, or galaxies. In these methods, one can calculate the statistical quantities by counting the number of objects. The relation between these quantities and the $n$-point correlation functions shows that they depend on all order correlations. So, they contain more information than the two-point correlation function, especially at small scales \citep{white1979hierarchy, bernardeau1992gravity, balian1989scale, szapudi1993higher}.\\
The number of objects in a specific volume such as a sphere is called counts-in-cells (CIC) \citep{1936rene.book.....H, 10.2307/40673267, balian1989scale}. We know that in small scales, the distribution of CIC or equivalently one-point probability of the matter density field is not Gaussian and follows approximately a log-normal distribution  \citep{10.1093/mnras/248.1.1, 10.1093/mnras/280.3.754}. The nearly log-normal behaviour of the cosmic matter density field is a feature of evolving perturbations from the linear Gaussian inflationary initial conditions \citep{guth1981inflationary, linde1982new} to the late time non-linear and non-Gaussian distribution. Accordingly, it is informative to investigate the shape of the probability distribution of the matter density field and its evolution\citep{10.1111/j.1365-2966.2008.13038.x, Bernardeau:2013dua, 10.1093/mnras/stw1074, Ivanov:2018lcg, Repp:2020etr, 10.1093/mnras/staa2073}. It is worth mentioning that other interesting ideas such as an Edgeworth expansion, skewness, and kurtosis analysis are argued in this field of study \citep{Colombi:1994sj, gaztanaga2000gravitational, shin2017new, Klypin:2017jjg, einasto2020evolution}. On large scales, the CIC contains the same information as the two-point correlation function. But there is more information encoded in the log-normal shape of the CIC distribution on small scale. Accordingly, the cosmological dependence of the CIC statistics is interesting to be explored in mildly and strongly non-linear regimes from an observational and theoretical point of view. \cite{uhlemann2020fisher} showed that the measured CIC statistics are sensitive to the cosmological parameters and neutrino mass. \cite{Repp:2020kfd} used the CIC to break the degeneracy between $\sigma_8$ and galaxy bias. Also, it is employed to study the primordial non-Gaussianity in LSS by measuring the ${f}_{\text{NL}}$ parameter \citep{Friedrich:2019byw}. Recently, \cite{Jamieson:2020wxf} introduced an approach to use a position-dependent one-point distribution of matter density as a cosmological observable. Besides all this progress, there are some limitations.
The results depend on the size and shape of the cell, where for large smoothing scales the distribution is Gaussian whereas, for small scales, the distribution is approximately log-normal. So for each cell size, one should calculate the CIC distribution, which is a computationally expensive procedure.
Finally, we conclude that although the CIC is a computationally costly process, it is advantageous to explore different regimes and compare the linear and non-linear scales. \\
Another quantity defined to study non-linear scales is the void probability (VP) function. It is defined as the probability of a volume (such as a sphere) to be empty. \cite{white1979hierarchy} derived the relation of the VP function with higher-order correlations and showed that the VP is a generating function for the CIC distribution. Moreover, a remarkable feature of the VP function is its scaling relation. It was shown that for some proposed phenomenological models of $n$-point correlation functions in a strongly non-linear regime, the VP function inherits a scaling relation. This relation has a universal behaviour as a function of scale \citep{white1979hierarchy, balian1989scale, bernardeau1992gravity, szapudi1993higher, munshi1999scaling}. Investigating the galaxy surveys and numerical simulations data exhibit the existence of this universal behaviour for galaxy magnitude and halo mass respectively. \citep{1986ApJ...306..358F, maurogordato1987void, 1991ApJ...369...30M,Croton:2004ac, Fry:2013pma}.

The VP function has been widely used in cosmological studies.
\cite{10.1093/mnras/195.4.857} used the VP function in large scales to study the $n$-point correlation function. \cite{Walsh:2019luq} used it to study the galaxy assembly bias. \cite{Paranjape:2020wuc} introduced the Voronoi volume function and its sensitivity to the cosmological parameters. 
\cite{Banerjee:2020umh,banerjee2021cosmological} introduced a new probe named the k-nearest neighbour (k-NN) distance from a Poisson distributed random points, which improves the constraints on the cosmological parameters. 
In this work, we use the spherical contact cumulative/probability distribution function (SC-CDF/PDF) which is complementary to the VP function. {{We should note that the SC-CDF is a known function in statistics \citep{chiu2013stochastic}, and as we are going to discuss in the next section, it can be calculated using the VP function. Also the reason of using this terminology will be pointed out later.}} \\
According to the definition of the VP function, it probes mainly the cosmic voids, and is not sensitive to the details of the structures. Accordingly, changing the halo finder method does not have a significant impact on the VP function. The nearest neighbour cumulative/probability distribution function (NN-CDF/PDF), on the other hand, is a probe that is more sensitive to the details of the structures \citep{clark1954distance,soneira1977there, martinez2001statistics, baddeley2006stochastic, chiu2013stochastic}. The NN-CDF is the complementary probability of finding an empty spherical region with a halo or galaxy at its centre.
So it is more sensitive to clustering and entails different information compared to the VP function. In this direction, by combining the VP function and the NN-CDF, we used a quantity called the J-function \citep{van1996nonparametric, Kerscher:1999gt} (see theoretical background section).\\
The relation of the previously mentioned probes to the cosmological parameters and the initial conditions of the universe is not straightforward, and we often need to perform a numerical simulation to obtain them. 
However, we can use these probes to compare different samples. These samples are made by considering different redshift or mass/magnitude limit of halos/galaxies. Both possibilities when one compares different samples are interesting. If the functions are mass-scale/redshift independent, it leads us to define universal scales. If the functions evolve with redshift or/and are mass scale-dependent, it could be a sign of an interesting physics behind this evolution/dependence. 

Finding the origin of the universality behaviour is crucial. It could be the result of gravitational instability and non-linear evolution or it could be a characteristic of the initial conditions.
Whatever the answer is, we can check the universality in different theories (simulations) and observations.\\
Accordingly, we prepare different samples, by considering mass limit and different redshift snapshots in simulations. We explore numerical N-body simulations as well as the galaxy group catalogues. We calculate the SC-CDF, NN-CDF, and J-function. Moreover, we study the SC-PDF, NN-PDF and calculate their moments. We also introduce the logarithmic moments to distinguish different samples. 
In this work, we show that the SC-CDF and NN-CDF probe non-linear scales differently. We employ the N-body simulations in the redshift range of zero to unity with different mass limits ($10^{11}-10^{14}M_{\odot}h$) to study the time evolution and mass dependence of the SC-CDF and NN-CDF. {{Furthermore, we consider the mock galaxy catalogues as a better approximation to the observational data. We study the mock central galaxies, where we set conditions on the magnitude to compute the NN-CDF and SC-CDF statistics.}}

The structure of this work is as follows: In Section~\ref{Sec2}, we review the theoretical background of the non-linear quantities, the SC-CDF, NN-CDF and $J$-function. In Section~\ref{Sec3} we go through the numerical simulations data and calculate these measures for different samples. {{{In Section~\ref{Sec4} we show results for the galaxy groups and the mock central galaxies. Finally, in Section~\ref{Sec5} we conclude and point out the future remarks.}}}
\section{Theoretical background}
\label{Sec2}
In this section, we use a class of summary statistics that encapsulate information from higher-order correlations. We consider the SC-CDF (which is equivalent to the VP function), NN-CDF, and J-function. There are three theoretical points of view to study these quantities; a) Their relation to $n$-point correlation function (see Subsection~\ref{Sub1Sec2}). b) The shape of their probability distributions. In this direction, we introduce the normal and log-normal distribution functions, which are relevant to the linear and non-linear statistics, respectively (see Subsection~\ref{Sub2Sec2}) and c) The scaling relations corresponding to these quantities (see Appendix~\ref{App_Uni}).
\\
\subsection{Relations to the $n$-point correlation functions} \label{Sub1Sec2}
The frequently used statistical measure for point processes \citep{chiu2013stochastic} is the two-point correlation function $\xi_2(r)$,
\begin{equation} \label{Eq: corr}
\langle \text{n}(\textbf{r}_1) \text{n}(\textbf{r}_1+\textbf{r})\rangle = [1+\xi_2(r)] n^2,
\end{equation}
where $\text{n}(\textbf{r})$ is the number density of points and $n = \langle \text{n}(\textbf{r}) \rangle$ is mean number density. In linear scales the density field is a Gaussian distribution and the two-point correlation $\xi_2(r)$ and $n$ have all the field information. In non-linear scales due to the non-Gaussianity we have to consider other $n$-point correlations $\xi_n(\textbf{r}_1,...,\textbf{r}_n)$. \cite{white1979hierarchy} introduced conditional correlation functions ($\Xi_i$) as a connection between the VP and $n$-point correlation functions.
\begin{align} \label{Eq: conditional corr}
\Xi_i(\textbf{r}_1,...,\textbf{r}_i;V) =& \sum_{j=0}^{\infty} \dfrac{(-n)^j}{j!} \\ \nonumber
& \int ...\int \xi_{i+j}(\textbf{r}_1,...,\textbf{r}_{i+j})\,dV_{i+1}...dV_{i+j},
\end{align}
where the integrals are taken over the volume $V$ and by definition $\xi_0 = 0$ and $\xi_1 = 1$. The ordinary correlation $\xi_n(\textbf{r}_1,...,\textbf{r}_n)$ corresponds to the probability that $n$-points defined in positions $\textbf{r}_1$, $\textbf{r}_2$ ... , $\textbf{r}_n$. Due to statistical homogeneity and isotropy, the correlations depend only on relative positions of the points. However, the conditional correlation have an additional property that the volume $V$ to be empty. So it depends on the position of points in the volume $V$\citep{white1979hierarchy}.\\
The SC-CDF is the cumulative distribution of the distance between randomly distributed positions in space and their nearest neighbour, taken from a sample. We show the SC-CDF by $F(r)$ where $r$ is the relative distance between random positions and their nearest neighbour. In other words, $F(r)$ is the probability that the nearest neighbour distance of a random point in space to be equal or smaller than $r$. 
On the other hand, the NN-CDF is the cumulative distribution of the distance between a halo/galaxy and their nearest neighbour (which itself is a halo/galaxy), both from the same sample. We show this function by $G(r)$. Equivalently, the $G(r)$ is the probability that the nearest neighbour distance of a halo/galaxy to be equal or smaller than $r$.

The complementary cumulative distribution function related to the SC-CDF is equal to the probability that all sample points to be out of the sphere with volume $V$. So there is a direct relation between the SC-CDF and the VP function (the probably that a sphere to be empty),
\begin{equation} \label{Eq: VP SCF}
P_0(V(r)) = 1 - F(r),
\end{equation}
where $P_0$ is the VP function and $V(r) = \dfrac{4\pi}{3}r^3$. \cite{white1979hierarchy} calculated the relation between the VP function and the zero conditional correlation function 
\begin{equation} \label{Eq: VP}
P_0(V) = \exp\left[ \Xi_0(V(r))\right]. 
\end{equation}
Equations~(\ref{Eq: conditional corr},\ref{Eq: VP}) show that all higher-order correlation functions influence the VP function and the SC-CDF.
Accordingly the SC-CDF is
\begin{equation}\label{Eq: SCF}
F(r) = 1 - \exp\left[ \Xi_0(V(r))\right].
\end{equation}
The complementary cumulative distribution $1-G(r)$, related to the NN-CDF $G(r)$ is the probability that a point to be in the centre of an empty sphere.
Using the conditional correlations we have
\begin{equation}\label{Eq: NND}
G(r) = 1 - \Xi_1(\textbf{r}_o;V(r)) \exp\left[ \Xi_0(V(r))\right],
\end{equation}
where $\textbf{r}_o$ is the position vector of the sphere's centre, which can be set to zero by a coordinate transformation. $\Xi_1(\textbf{r}_o,V(r))$ is the conditional probability of a point in the centre of an empty volume $V$ and $\exp\left[ \Xi_0(V(r))\right]$ is the probability of the void function.
Comparing the SC-CDF with the NN-CDF, we should note that the former depends only on the zeroth-order conditional correlation. However, the latter is related to the first conditional probability. It is important to notice that the NN-CDF is more sensitive to clustering compared to the SC-CDF. As voids occupy the larger portions of the cosmic web, the SC-CDF (which is the probability of a region to be empty) explores the larger scales. In contrast, the NN-CDF is useful to study the high-density regions and, accordingly, the smaller scales. {{Now it is more clear why we use the spherical contact terminology instead of the void probability. The SC-CDF, by definition, is similar to the NN-CDF, while the VP function is similar to complimentary of the NN-CDF. This is clear from equations~(\ref{Eq: SCF},\ref{Eq: NND}).\\ 
We can define the J-function as}}
\begin{equation} \label{Eq: J}
J(r) = \dfrac{1-G(r)}{1-F(r)} = \Xi_1(\textbf{r}_o;V(r)),
\end{equation}
which only has $\Xi_1(\textbf{r}_o;V(r))$ dependency. If the data is a Poisson random set with zero correlation between points, then we have
\begin{equation} \label{Eq: Poisson}
F(r) = G(r) = 1 - \exp\left(-\dfrac{4\pi}{3} n r^3\right).
\end{equation}
Thus for a sample without any correlation, we have $J=1$. Accordingly, the J-function is a better measure of clustering compared to the SC-CDF and NN-CDF. The J-function for the samples with positive correlation is below unity $J<1$, and for anti-correlated ones, we have $J>1$, which is helpful to distinguish these two types as well \citep{Kerscher:1999gt}. \\
Moreover, we can find a relation for the probability distribution functions by taking the derivative of equation~(\ref{Eq: SCF}) and equation~(\ref{Eq: NND}) with respect to $r$. For this task, first we have to calculate the derivative of the conditional correlations, equation~(\ref{Eq: conditional corr}) with respect to the variable $r$, radius of the sphere $V$. Due to spherical symmetry, we have 
\begin{equation} \label{Eq: dW0}
    \dfrac{d}{dr}\Xi_0(V(r)) = (-4\pi r^2 n) \Xi_1(\textbf{r}_s;V(r)),
\end{equation}
where $\textbf{r}_s$ is a vector on the boundary of sphere and $\Xi_1(\textbf{r}_s;V(r))$ is only a function of $|\textbf{r}_s| = r$.
Using equations~(\ref{Eq: SCF},\ref{Eq: dW0}), we have
\begin{equation}\label{Eq: SCF PDF}
\left| \dfrac{dF(r)}{dr} \right| = 4\pi r^2 n \, \Xi_1(\textbf{r}_s;V) \exp\left[\Xi_0(V)\right].
\end{equation}
To calculate the NN-PDF we must know the derivative of the first conditional correlation at position $\textbf{r}_o$. Again due to spherical symmetry we can calculate the derivatives analytically. From equation~(\ref{Eq: conditional corr}) we find that
\begin{equation}
    \dfrac{d}{dr}\Xi_1(\textbf{r}_o;V(r)) = (-4\pi r^2 n) \, \Xi_2(\textbf{r}_s,\textbf{r}_o;V(r)),
\end{equation}
where $\textbf{r}_o$ is the centre of the sphere, and we can set $\textbf{r}_o=0$ by a coordinate transformation. So $\Xi_2(\textbf{r}_s,\textbf{r}_o;V(r))$ is symmetric and only a function $r$. Finally, we have
\begin{align}\label{Eq: NND PDF}
\left| \dfrac{dG(r)}{dr} \right|= 4\pi r^2 n\left[ \Xi_1(\textbf{r}_s;V)\, \Xi_1(\textbf{r}_o;V) + \Xi_2(\textbf{r}_s,\textbf{r}_o;V)\right] \exp\left[\Xi_0(V)\right] ,
\end{align}
where again $\textbf{r}_s$ is a vector on the sphere's boundary. By comparing the SC-PDF and the NN-PDF from equations~(\ref{Eq: SCF PDF},\ref{Eq: NND PDF}), we find that the NN-PDF is related to the higher-order conditional probabilities in contrast to the SC-PDF. We also can compare the SC-PDF with the NN-CDF from equation~(\ref{Eq: NND}). The two equations have similarities and both depend on $\Xi_1$. However, the SC-PDF is the probability that depends on $\textbf{r}_s$, while the NN-CDF depends on $\textbf{r}_0$.\\
One way to calculate the SC-CDF for a data set is to draw spheres with radius $r$ in space randomly. To find the SC-CDF we calculate the ratio of the empty to total spheres as 
\begin{equation} \label{Eq: practical}
1-F(r) = \dfrac{N_{\text{empty}}}{N_{\text{total}}}.
\end{equation}
This approach works for the NN-CDF too.  We set the centre of all spheres on the sample points and repeat the procedure in different $r$.
The efficient approach to calculate the SC-CDF/SC-PDF is to create a randomly distributed Poisson sample.  We calculate the distance between random points to their nearest neighbour chosen from observation/simulation data points. We can directly calculate the probability and cumulative distributions with ordered distances. For the NN-CDF/NN-PDF the procedure is similar, but we choose all the points from observation/simulation data. For the SC-CDF the number of Poisson sample points is arbitrary. 
We calculate the SC-CDF/SC-PDF with the same number of random points as the observation/simulation data. We follow this procedure to have a better comparison with NN-CDF/NN-PDF. Accordingly, we obtain the $J-$ function consistently (for another choice of random point's number, see \cite{Banerjee:2020umh}).
Hereafter, we calculate the aforementioned functions with respect to a new introduced dimensionless variable $x$ defined as
\begin{equation} \label{Eq: x}
    x = \left(\dfrac{4\pi}{3} n\right)^{1/3} r.
\end{equation}
The privilege of using the dimensionless variable $x$ is that we omit the effect of the mean number density. Accordingly, the SC-NN functions depend only on the $n$-point correlation functions.
Further, we can use $x$ in the conditional correlations as well, equation~(\ref{Eq: conditional corr}), to remove the effect of number density. For more details see, Appendix~(\ref{App_Uni}). 

\subsection{Probability distributions} \label{Sub2Sec2}
The normal (Gaussian) probability distribution function $\mathcal{N}(\mu,\sigma)$ is defined as follows,
\begin{equation} \label{Eq: normal}
f_n(x) = \dfrac{1}{\sqrt{2\pi} \sigma} \exp[-(x-\mu)/2\sigma^2].
\end{equation}
In linear scales, the matter distribution is normal as a result of the single field inflationary models. In non-linear scales the matter distribution is nearly log-normal distribution as below 
\begin{equation} \label{Eq: lognormal}
f_l(y) = \dfrac{1}{\sqrt{2\pi} y \sigma} \exp[-(\log(y)-\mu)/2\sigma^2].
\end{equation}
where the above equation is obtained by changing variable $y\equiv e^x$ in a normal distribution, equation~(\ref{Eq: normal}).
The difference between the normal and log-normal distribution from the central limit theorem point of view is discussed in \cite{10.1093/mnras/248.1.1}, as well. The theorem works for a stochastic variable, which is a sum over other variables $X = \sum_{i=1}^N X_i$. For large $N$ statistics the distribution tends to be normal  $f(x) \rightarrow \mathcal{N}(\mu,\sigma)$. In our context, in linear scales, the matter density contrast is a sum over independent modes of Fourier space, consequently its statistics is Gaussian.
For non-linear quantities where $Y = \Pi_{i=1}^N X_i$ and for large number $N$, the logarithm of $Y$ can be written as $\log(Y) = \sum_{i=1}^N \log(X_i)$. So the central limit theorem implies a normal distribution, equation~(\ref{Eq: lognormal}), for $\log(Y)$ and log-normal distribution for $Y$. In standard perturbation theory as a result of mode coupling the non-linear density contrast is constructed by multiplication of linear density contrast with corresponding kernels \citep{Bernardeau:2001qr} and as a result we expect an approximately log-normal distribution for matter density in non-linear scales.
 
In the next section,  we calculate the moments of the SC-PDF, where the skewness is not small. We introduce the skew normal distribution as
\begin{equation}\label{Eq: Skewnormal}
f_s(x) = f_n(x) [1+\text{erf}(\alpha\bar{x})],
\end{equation}
where $f_n(x)$ is the normal distribution (equation~(\ref{Eq: normal})), $\alpha$ is a constant, $\bar{x}=\dfrac{x-\mu}{\sqrt{2}\sigma}$ and $\text{erf}$ is the error function,
\begin{equation} \label{Eq: errorfunction}
\text{erf}(x) = \dfrac{2}{\sqrt{\pi}} \int_0^x e^{-t^2} \, dt.
\end{equation}
In this work, we fit the SC-PDF and NN-PDF with the mentioned distributions and discuss  the probable deviations. 
We can also characterise the distribution functions  by their moments. {{The $n$th central moment of a distribution function $f(x)$ is defined as,
\begin{equation} \label{Eq: moments}
\mu_n = \int (x-\mu)^n \, f(x) \, dx,
\end{equation}
where $\mu$ is the mean of distribution. The zeroth central moment is unity $\mu_0 = 1$ for a normalised distribution and we have $\mu_1 = 0$ by definition. We define mean and variance of the distribution by $s_1=\mu$ and $s_2=\sqrt{\mu_2}$. We also define the skewness and kurtosis by $\tilde{s}_3 = \mu_3/s_2^3$ and $\tilde{s}_4 = \mu_4/s_2^4$, respectively. For a normal distribution equation~(\ref{Eq: normal}), variance $s_2 = \sigma$, skewnees is $\tilde{s}_3 = 0$ and kurtosis is $\tilde{s}_4 = 3$. In this work, we simply refer to $s_1$, $s_2$, $\tilde{s}_3$ and $\tilde{s}_4$ as moments of a distribution. We extent this concept to the logarithmic case and introduce the corresponding central moments as
\begin{equation} \label{Eq: logmoments}
\kappa_n = \int (\ln(x)-\mu')^n \, f(x) \, dx,
\end{equation}
where $\mu'$ is the logarithmic mean of the distribution. Note that for a normalised distribution $\kappa_0 = 1$ and $\kappa_1 = 0$. For a log-normal distribution equation~(\ref{Eq: lognormal}), we obtain the higher moments as  $\kappa_2 = \sigma^2$, $\kappa_3 = 0$, and $\kappa_4 = 3 \sigma^4$. We define the logarithmic mean and variance of distribution by $l_1=\mu'$ and $l_2=\sqrt{\kappa_2}$. 
Accordingly, the logarithmic skewness and kurtosis of a log-normal distribution are $\tilde{l}_3 = \kappa_3/l_2^3 = 0$, and $\tilde{l}_4 = \kappa_4/l_2^4 = 3$. The logarithmic moments are especially useful when we study the  distributions in non-linear scales. 
In this work we refer to $l_1$, $l_2$, $\tilde{l}_3$ and $\tilde{l}_4$ as logarithmic moments of a distribution.}} \\
The normal and log-normal moments naturally define two classes of distributions. As a specific indicator, if the kurtosis of distribution is $\tilde{s}_4 \sim 3$, the distribution is nearly linear or normal and if the logarithmic distribution has a kurtosis  $\tilde{l}_4 \sim 3$ the distribution is nearly non-linear or log-normal. In the next section, we will show that the SC-PDF is linear and, the NN-PDF is a non-linear distribution.
\section{Simulations} \label{Sec3}
As we discussed, it is intricate to calculate the NN-CDF/PDF, SC-CDF/PDF, and J-functions by analytical prescription of Section~\ref{Sec2}. Accordingly, we explore these functions in cosmological simulations. In the first subsection, we explain the simulations and methods we used in this work, and we will present the results in the second subsection.
\subsection{SMDPL simulation} \label{Sec3: details}
\begin{table}
\centering
\caption{We present the details of the different dark matter halo samples from the SMDPL simulation. The conditions on mass and redshift are shown in columns 1 and 2, respectively. And $r_{\star} = 1/\sqrt[3]{(4\pi n/3)}$ is represented in column 3.}
\label{Ta: DM samples}
\begin{tabular}{ccc}
\hline
$\log(M_{\lim}/M_{\odot}h)$ & redshift(z) & $r_{\star}$ (Mpc/h) \\
\hline
\hline
11 & 0 & 1.963\\
11 & 0.5 & 1.942\\
11 & 1 & 1.943\\
\hline
11.5 & 0 & 2.759\\
11.5 & 0.5 & 2.758\\
11.5 & 1 & 2.803\\
\hline
12 & 0 & 3.884\\
12 & 0.5 & 3.949\\
12 & 1 & 4.114\\
\hline
12.5 & 0 & 5.530\\
12.5 & 0.5 & 5.760\\
12.5 & 1 & 6.234\\
\hline
13 & 0 & 8.019\\
13 & 0.5 & 8.689\\
13 & 1 & 9.981\\
\hline
13.5 & 0 & 12.083\\
13.5 & 0.5 & 13.953\\
13.5 & 1 & 17.624\\
\hline
14 & 0 & 19.781\\
14 & 0.5 & 25.163\\
14 & 1 & 36.904\\
\hline
\end{tabular}
\end{table}
In this work, we use the simulations publicly available in the "MultiDark" project \citep{riebe2013multidark}. The project includes several simulations with different box sizes, seed numbers, and different number of particles\footnote{ Find the complementary information of these simulations in the web page {https://www.cosmosim.org}}. As we are interested in non-linear scales, roughly below 10 Mpc, the resolution should be high enough to capture the relevant scales. To have a statistically enough number of halos, we need a large box size to probe the end tails of the distributions. To satisfy these two conditions together, we choose the Small MultiDark Planck (SMDPL) simulation. The SMDPL box size is $400 \text{Mpc/h}$ and it has $3840^3$ particles. The cosmological parameters are as Planck 2014 \citep{Ade:2013zuv}, $ h= 0.6777$, $\Omega_{\Lambda} = 0.692885$, $\Omega_{m}=0.307115$, $\Omega_b=0.048206$, $n_s=0.96$ and $\sigma_8 = 0.8228$. Three redshifts $z=0$, $z=0.49$ \footnote{ In figures and, tables we show it with one digit accuracy $z=0.5$}, and $z=1$ are considered in simulations to explore the redshift evolution of the quantities. For each redshift, we create different samples by limiting the mass of halos to find the scale dependence of the quantities. Accordingly, we choose $7$ different mass limits $\log(M/M_{\odot}h)>\{11.5,12,12.5,13,13.5,14\}$. In total, we have $3\times 7=21$ different samples, which we use to calculate the mass and redshift dependence of these quantities (i.e., the SC-CDF, NN-CDF, and J-function). In Table~(\ref{Ta: DM samples})  we summarize the characteristics of the different samples, and report the $r_{\star} = 1/\sqrt[3]{(4\pi n/3)}$ for each sample. Note that throughout this paper, all the statistical quantities including moments are expressed with the dimensionless variable $x = r/r_{\star}$ in which $r$ is the comoving length in $\text{Mpc}/h$ (Section~\ref{Sec2}). We explore the results from the two other simulations of the MultiDark project to check the consistency and robustness of the result concerning simulation characteristics in Appendix~(\ref{APP_Sim}).\\
The statistical quantities we introduced in the previous section could be computed using dark matter halos or particles in N-body simulations. In this work, we use the position of dark matter halos to measure these quantities.
We use dark matter halos provided by \cite{riebe2013multidark}. These halos have been identified using the FoF method (\cite{lacey1994merger}) and the linking-length is set to $b=0.2$. As we have the positions and properties of the dark matter halos, we can calculate the SC-CDF, NN-CDF, and J-function for these halos directly using the methods introduced in Section~\ref{Sec2}. {{In this work, we do not consider periodic boundary conditions. However, as we are going to discuss in details in Appendix~\ref{APP_Sim} the effect is negligible and thus can be ignored. }}\\
We calculate the statistical errors by dividing the simulation box into 27 similar sub-boxes, and use them as sub-samples with the assumption of the statistical homogeneity. We calculate the quantities in each sub-sample, and we report the mean and standard deviation of these sub-boxes with corresponding conservative error bars. For more details see Appendix~\ref{APP_Sim}. 
We also follow a similar approach to find errors in observational data in Section~\ref{Sec4}.
\begin{figure}
\centering
\includegraphics[scale=0.3]{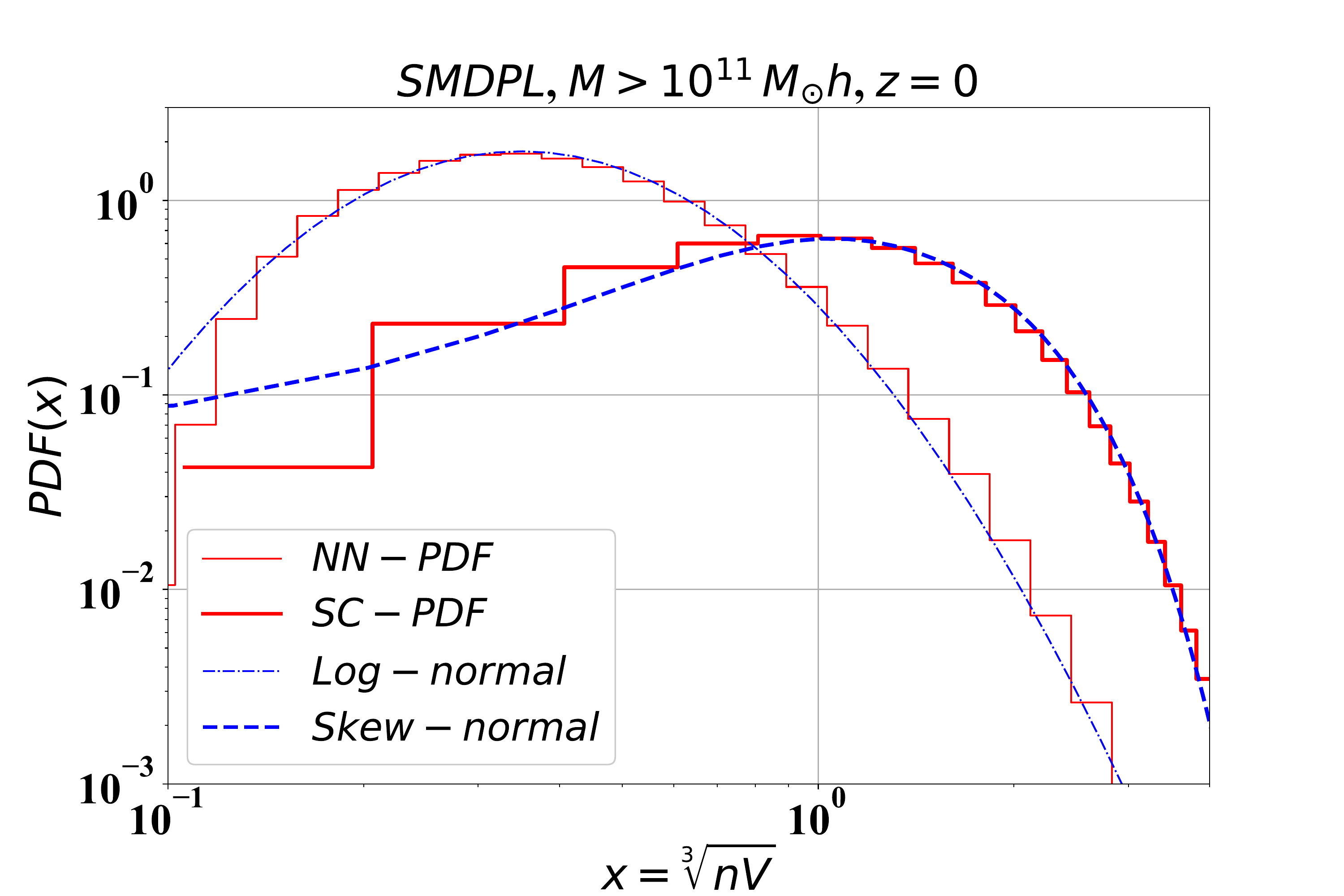}
\caption{The NN-PDF and  SC-PDF for dark matter halos are plotted as a function of the dimensionless variable $x$ for the SMDPL simulation. In this figure, we set the mass limit to $M > 10^{11} M_{\odot}h$ and the redshift is $z=0$. Thin red line refers to the NN-PDF and the thick red line refers to the SC-PDF. Dot-dashed thin blue line is the theoretical log-normal curve equation~(\ref{Eq: lognormal}) with $\mu =-0.75,\sigma=0.55$.The dashed thick blue line is the skew-normal curve equation~(\ref{Eq: Skewnormal}) with $\mu =0.55,\sigma=1,\alpha=2.6$.}
\label{fig:Hist}
\end{figure}
\begin{figure}
\centering
\includegraphics[width=0.43\textwidth]{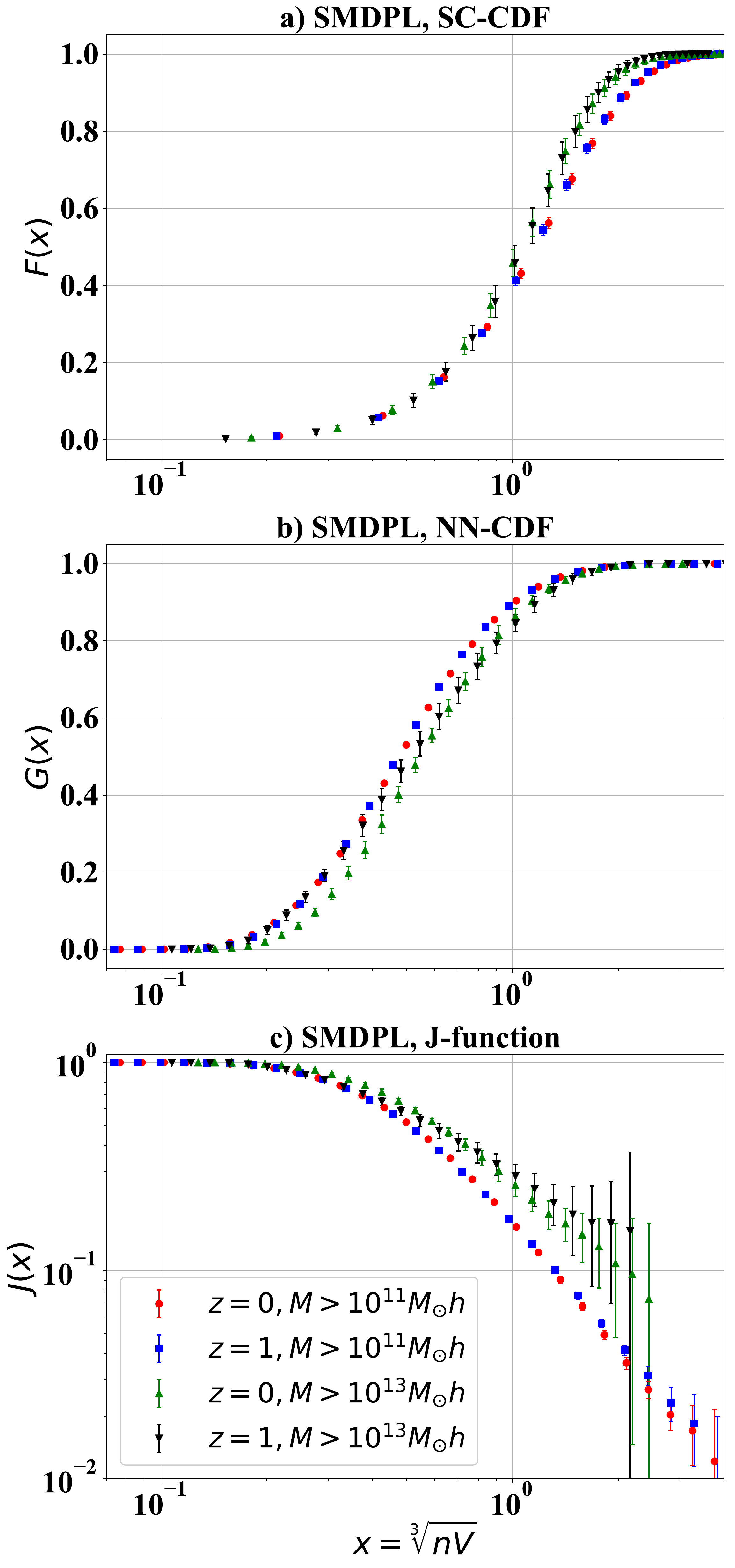}
\caption{In this figure, a) $F(x)$, the SC-CDF, b) $G(x)$, the NN-CFD and c) $J(x)=\dfrac{1-G(x)}{1-F(x)}$ as a function of  $x=\sqrt[3]{nV}$ are plotted using halos from the SMDPL simulation. The red circles ($z=0$) and the blue squares ($z=1$) correspond to lower mass limit of $10^{11}M_{\odot}h$ and green up-pointing triangles ($z=0$) and black down-pointing triangles ($z=1$) correspond to lower mass limit of $10^{13}M_{\odot}h$.}
\label{fig:CDF}
\end{figure}
\subsection{Results}
We discuss the mass scale and redshift dependence of the SC-CDF/PDF, NN-CDF/PDF, and J-function using cosmological $\text{N}$-body simulations. Specifically, in this subsection, we represent results obtained from the dark matter halos of the SMDPL simulation.
To see the general behaviour of the SC-PDF and NN-PDF in Fig.~\ref{fig:Hist}, we plot the SC-PDF (thick solid red line) and the NN-PDF (thin solid red line) of dark matter halos as a function of $x=\sqrt[3]{nV}$. 
In Fig.~\ref{fig:Hist}, the redshift is $z=0$ and the mass limit is set to $M > 10^{11} M_{\odot}$. It is clear that both the SC-PDF and  NN-PDF have visible peaks around $x\sim 0.5$ and $x \sim 1.2$ and these numbers can be translated to a comoving length of $r=r_{\star} x \approx 1$ Mpc/h and $r \approx 2.3$ Mpc/h respectively. Note that for this specific sample $r_{\star}=1.963$ (see Table~(\ref{Ta: DM samples})).
Based on Fig.~\ref{fig:Hist}, the SC-PDF approximately follows skew-normal, equation~(\ref{Eq: Skewnormal}), with parameters $\mu =0.55,\sigma=1,\alpha=2.6$ (dashed thick blue line), and the NN-PDF approximately is log-normal, equation~(\ref{Eq: lognormal}), with parameters $\mu =-0.75,\sigma=0.55$ (dotted-dashed thin blue line). Here, we have the idea of using the normal and log-normal distributions as the anchor of our study for linear and non-linear scales.
It is worth mentioning that the theoretical curves are not precise, and the simulation data deviates from these curves, especially at the tail of the distributions.
For example the distance variable $x$, in equation~(\ref{Eq: x}), is always positive, while our fitting (normal and the skew-normal) distributions would give non-zero  (although small) values for negative $x$ which is not physical.
This issue also exists when considering the matter density field as a normal distribution \cite{gaztanaga2000gravitational}. However, the log-normal distribution seems to be more appropriate to address this issue when we use positive variables.
Testing the robustness of more complex distributions (for example, the Edgeworth expansion) will be the subject of our future works.\citep{gaztanaga2000gravitational,shin2017new}
Considering the general behaviour of the distributions, the SC-PDF is nearly skew-normal, and the NN-PDF is nearly log-normal. The SC-PDF is defined based on the distance between one halo to a random point in a sample, while the NN-PDF is defined according to the position between two halos. As we mentioned previously, due to the central limit theorem, the SC-PDF is a linear distribution (it is related to the general shape of the cosmic web) and the NN-PDF is a non-linear distribution (it depends on details of the structures' distribution).\\
We make four samples from Table~(\ref{Ta: DM samples}). This is done by applying conditions on mass scale, $\log(M/M_{\odot} h)> \{11,13\}$, and redshift $z=\{0,1\}$. In Fig.~\ref{fig:CDF}, we present a) the SC-CDF, b) the NN-CDF, and c) the J-function in terms of $x=\sqrt[3]{nV}$. The red circles represent $z=0$ with lower mass limit of $10^{11}M_{\odot}h$, the blue squares correspond to $z=1$ with lower mass limit of $10^{11}M_{\odot}h$, green up-pointing triangles correspond to $z=0$ with lower mass limit of $10^{13}M_{\odot}h$ and black down-pointing triangles correspond to $z=1$ with lower mass limit of $10^{13}M_{\odot}h$. \\
According to part (a) of Fig.~\ref{fig:CDF}, the SC-CDF is redshift-independent while it shows a mass-limit dependence in the range of $x>1$.
\citep{Banerjee:2020umh} asserts that when we study tail of the distributions with a new quantity, namely "peaked CDF", we can distinguish the SC-CDF in two different redshift samples.
In panel (b) of Fig.~\ref{fig:CDF}, the difference between distinct mass scales is clear (similar to the panel a). Moreover, the NN-CDF  has a clear redshift evolution in the redshift range of zero to unity for the lower mass limit of $10^{13}M_{\odot}h$.
Based on the curves in Fig.~\ref{fig:CDF} the SC-CDF is a better function to find the similarities, while the NN-CDF helps to distinguish different samples. 
In panel (c) of Fig.~\ref{fig:CDF} we plot the J-function. The J-function is sensitive to the clustering, so we see that in the lower mass ranges, the halos are more clustered in the deep non-linear regime. \\
\begin{figure*}
\includegraphics[width=1\textwidth]{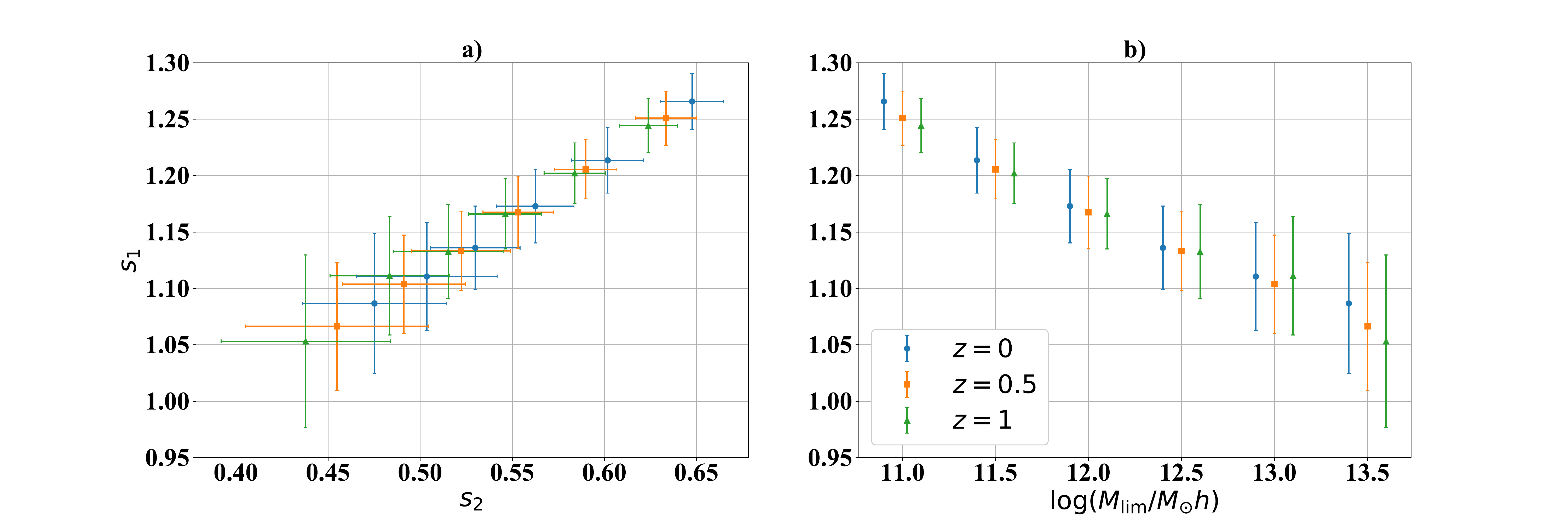}
\caption{The mean of the SC-PDF for dark matter halos of the SMDPL samples, $s_1$ defined in Section~\ref{Sub2Sec2}, is plotted for different mass scales with respect to a) $s_2$ and b) mass limits. Blue circles correspond to $z=0$, orange squares are for $z=0.5$, and green triangles are for $z=1$. In panel (b), the mass scales are shifted slightly for clarification.} \label{fig:s}
\end{figure*}
\begin{figure*}
\includegraphics[ width=1\textwidth]{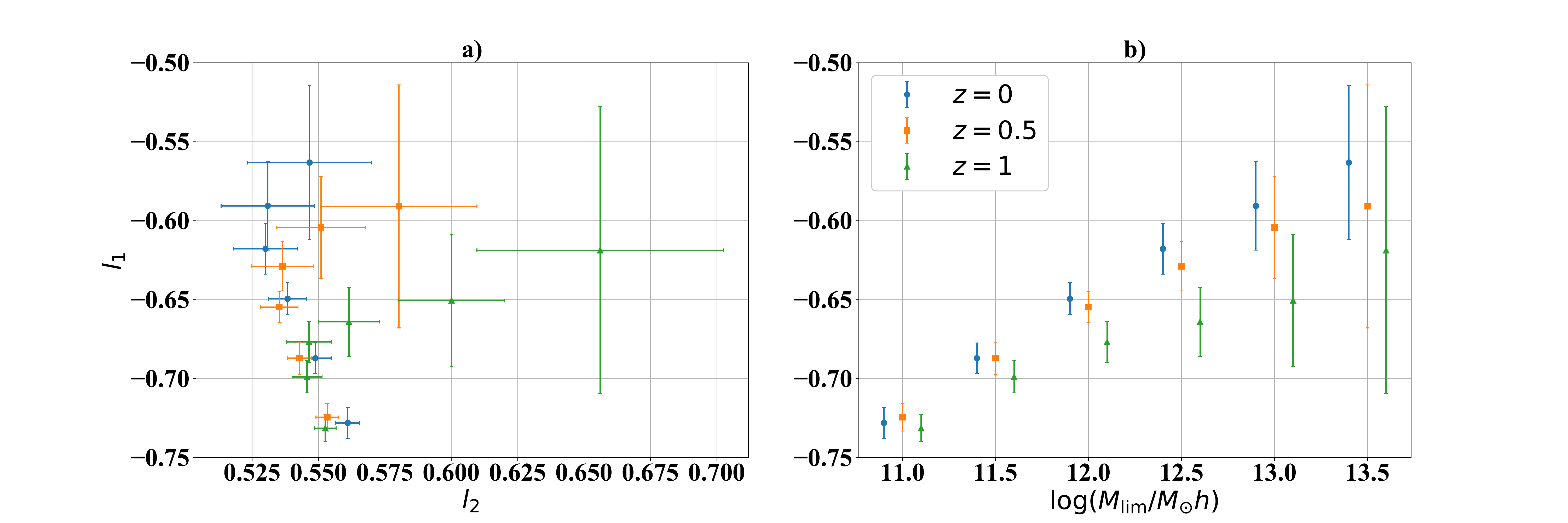}
\caption{The logarithmic mean of the NN-PDF for dark matter halos of the SMDPL samples, $l_1$ defined in Section~\ref{Sub2Sec2}, is plotted versus a) $l_2$ and b) mass limits for different mass scales. The labels are the same as Fig.\ref{fig:s}.  In panel (b), the mass scales shifted slightly for clarification.}
\label{fig:l}
\end{figure*}
\begin{table*}
\centering
\caption{The moments of the SC-PDF for the samples of Table~(\ref{Ta: DM samples}) calculated using equation~(\ref{Eq: moments}). The condition on the mass and redshift are presented in column 1 and 2 respectively. $s_1$, $s_2$, $\tilde{s}_3$ and $\tilde{s}_4$ are reported in the next columns, respectively.}
\label{Ta: SCF}
\begin{tabular}{cccccc}
\hline
$\log(M_{\lim}/M_{\odot}h)$ & redshift(z) & $s_1$ & $s_2$ & $\tilde{s}_3$ & $\tilde{s}_4$ \\
\hline
\hline
11 & 0 & $1.265 \pm 0.025$ & $0.647 \pm 0.017$ & $0.845 \pm 0.047$ & $3.827 \pm 0.192$ \\
11 & 0.5 & $1.251 \pm 0.024$ & $0.633 \pm 0.016$ & $0.830 \pm 0.045$ & $3.819 \pm 0.193$ \\
11 & 1 & $1.244 \pm 0.024$ & $0.624 \pm 0.016$ & $0.798 \pm 0.043$ & $3.711 \pm 0.177$ \\
\hline
11.5 & 0 & $1.215 \pm 0.027$ & $0.601 \pm 0.019$ & $0.780 \pm 0.055$ & $3.677 \pm 0.208$ \\
11.5 & 0.5 & $1.206 \pm 0.028$ & $0.591 \pm 0.018$ & $0.758 \pm 0.049$ & $3.622 \pm 0.210$ \\
11.5 & 1 & $1.204 \pm 0.027$ & $0.585 \pm 0.016$ & $0.737 \pm 0.051$ & $3.584 \pm 0.183$ \\
\hline
12 & 0 & $1.172 \pm 0.032$ & $0.563 \pm 0.022$ & $0.713 \pm 0.082$ & $3.520 \pm 0.304$ \\
12 & 0.5 & $1.168 \pm 0.030$ & $0.554 \pm 0.019$ & $0.685 \pm 0.052$ & $3.441 \pm 0.202$ \\
12 & 1 & $1.166 \pm 0.030$ & $0.545 \pm 0.017$ & $0.647 \pm 0.051$ & $3.353 \pm 0.182$ \\
\hline
12.5 & 0 & $1.136 \pm 0.037$ & $0.530 \pm 0.024$ & $0.667 \pm 0.097$ & $3.456 \pm 0.357$ \\
12.5 & 0.5 & $1.133 \pm 0.035$ & $0.522 \pm 0.027$ & $0.636 \pm 0.141$ & $3.419 \pm 0.538$ \\
12.5 & 1 & $1.133 \pm 0.042$ & $0.515 \pm 0.030$ & $0.603 \pm 0.158$ & $3.373 \pm 0.597$ \\
\hline
13 & 0 & $1.111 \pm 0.048$ & $0.504 \pm 0.038$ & $0.638 \pm 0.235$ & $3.508 \pm 0.805$ \\
13 & 0.5 & $1.104 \pm 0.044$ & $0.491 \pm 0.033$ & $0.614 \pm 0.181$ & $3.460 \pm 0.724$ \\
13 & 1 & $1.111 \pm 0.053$ & $0.483 \pm 0.032$ & $0.511 \pm 0.168$ & $3.173 \pm 0.505$ \\
\hline
13.5 & 0 & $1.087 \pm 0.062$ & $0.475 \pm 0.039$ & $0.545 \pm 0.182$ & $3.150 \pm 0.560$ \\
13.5 & 0.5 & $1.066 \pm 0.057$ & $0.455 \pm 0.050$ & $0.436 \pm 0.204$ & $2.992 \pm 0.569$ \\
13.5 & 1 & $1.053 \pm 0.076$ & $0.438 \pm 0.046$ & $0.495 \pm 0.315$ & $3.172 \pm 1.135$ \\
\hline
14 & 0 & $1.025 \pm 0.071$ & $0.427 \pm 0.060$ & $0.380 \pm 0.419$ & $2.987 \pm 1.317$ \\
14 & 0.5 & $1.032 \pm 0.092$ & $0.412 \pm 0.073$ & $0.236 \pm 0.344$ & $2.562 \pm 0.516$ \\
14 & 1 & $0.960 \pm 0.130$ & $0.353 \pm 0.116$ & $0.248 \pm 0.499$ & $2.444 \pm 0.613$ \\
\hline
\end{tabular}
\end{table*}
\begin{table*}
\centering
\caption{The logarithmic moments of the NN-PDF for the samples of Table~(\ref{Ta: DM samples}) calculated by equation~(\ref{Eq: logmoments}). The condition and limit on mass and redshift are presented in column 1 and 2 respectively. $l_1$, $l_2$, $\tilde{l}_3$ and $\tilde{l}_4$ are in following columns, respectively.}
\label{Ta: NND}
\begin{tabular}{cccccc}
\hline
$\log(M_{\lim}/M_{\odot}h)$ & redshift(z) & $l_1$ & $l_2$ & $\tilde{l}_3$ & $\tilde{l}_4$ \\
\hline
\hline
11 & 0 & $-0.728 \pm 0.010$ & $0.561 \pm 0.005$ & $0.120 \pm 0.011$ & $2.706 \pm 0.015$ \\
11 & 0.5 & $-0.725 \pm 0.009$ & $0.553 \pm 0.004$ & $0.157 \pm 0.010$ & $2.708 \pm 0.015$ \\
11 & 1 & $-0.731 \pm 0.009$ & $0.553 \pm 0.004$ & $0.199 \pm 0.010$ & $2.711 \pm 0.014$ \\
\hline
11.5 & 0 & $-0.687 \pm 0.010$ & $0.549 \pm 0.006$ & $0.099 \pm 0.015$ & $2.679 \pm 0.023$ \\
11.5 & 0.5 & $-0.687 \pm 0.010$ & $0.543 \pm 0.005$ & $0.146 \pm 0.016$ & $2.688 \pm 0.018$ \\
11.5 & 1 & $-0.699 \pm 0.010$ & $0.546 \pm 0.006$ & $0.193 \pm 0.021$ & $2.670 \pm 0.024$ \\
\hline
12 & 0 & $-0.649 \pm 0.010$ & $0.538 \pm 0.007$ & $0.076 \pm 0.027$ & $2.637 \pm 0.035$ \\
12 & 0.5 & $-0.655 \pm 0.010$ & $0.535 \pm 0.007$ & $0.139 \pm 0.025$ & $2.641 \pm 0.038$ \\
12 & 1 & $-0.677 \pm 0.013$ & $0.546 \pm 0.009$ & $0.191 \pm 0.028$ & $2.586 \pm 0.043$ \\
\hline
12.5 & 0 & $-0.618 \pm 0.016$ & $0.530 \pm 0.012$ & $0.081 \pm 0.041$ & $2.606 \pm 0.062$ \\
12.5 & 0.5 & $-0.629 \pm 0.016$ & $0.536 \pm 0.012$ & $0.151 \pm 0.047$ & $2.565 \pm 0.074$ \\
12.5 & 1 & $-0.664 \pm 0.022$ & $0.561 \pm 0.011$ & $0.167 \pm 0.056$ & $2.477 \pm 0.065$ \\
\hline
13 & 0 & $-0.591 \pm 0.028$ & $0.531 \pm 0.018$ & $0.102 \pm 0.045$ & $2.548 \pm 0.108$ \\
13 & 0.5 & $-0.604 \pm 0.032$ & $0.551 \pm 0.017$ & $0.140 \pm 0.077$ & $2.463 \pm 0.097$ \\
13 & 1 & $-0.651 \pm 0.042$ & $0.600 \pm 0.020$ & $0.078 \pm 0.085$ & $2.292 \pm 0.110$ \\
\hline
13.5 & 0 & $-0.563 \pm 0.049$ & $0.547 \pm 0.023$ & $0.064 \pm 0.125$ & $2.445 \pm 0.211$ \\
13.5 & 0.5 & $-0.591 \pm 0.077$ & $0.580 \pm 0.029$ & $-0.004 \pm 0.137$ & $2.309 \pm 0.190$ \\
13.5 & 1 & $-0.619 \pm 0.091$ & $0.656 \pm 0.046$ & $-0.105 \pm 0.242$ & $2.250 \pm 0.236$ \\
\hline
14 & 0 & $-0.508 \pm 0.074$ & $0.563 \pm 0.050$ & $-0.032 \pm 0.237$ & $2.330 \pm 0.348$ \\
14 & 0.5 & $-0.529 \pm 0.131$ & $0.641 \pm 0.082$ & $-0.263 \pm 0.277$ & $2.320 \pm 0.457$ \\
14 & 1 & $-0.511 \pm 0.270$ & $0.699 \pm 0.228$ & $-0.241 \pm 0.533$ & $1.932 \pm 0.497$ \\
\hline
\end{tabular}
\end{table*}
To study the similarities and differences of the samples more precisely, we calculate the moments of the SC-PDF and report it in Table~(\ref{Ta: SCF}). In Table~(\ref{Ta: NND}), we calculate the logarithmic moments of the NN-PDF. In general, calculating the moments helps us to summarise the information about the distributions of samples. The first and second columns of Table~(\ref{Ta: SCF}) show conditions on mass scale and redshift and the other four columns are $s_1$, $s_2$, $\tilde{s}_3$ and $\tilde{s}_4$ respectively ($l_1$, $l_2$, $\tilde{l}_3$ and $\tilde{l}_4$ for NN-PDF in Table~(\ref{Ta: NND})). Note that in Table~(\ref{Ta: SCF}) kurtosis, $\tilde{s}_4$, of the SC-PDF varies from ($3.8 \pm 0.2$) for mass scale $10^{11}$ to ($3\pm1$) for mass scale $10^{14}$. This again indicates that we can interpret the SC-PDF as a nearly normal distribution. The situation is different for the NN-PDF when logarithmic kurtosis $\tilde{l}_4$ is approximately 3. According to Table~(\ref{Ta: NND}) $\tilde{l}_4$ varies from ($2.70 \pm 0.02$) for mass scale $10^{11}$ to ($2.3\pm0.3$) for mass scale $10^{14}$. In Appendix~(\ref{App_Gen}) we study general behaviour of the SC-PDF and the NN-PDF in more details.
For the normal distribution, the skewness is zero, so the non-zero skewness shows the deviation from a normal distribution. Comparing Tables~(\ref{Ta: SCF},\ref{Ta: NND}), columns 5, the SC-PDF is more skewed than the NN-PDF. For the SC-PDF, the skewness is between $0.2$ and $0.8$, and the skewness is above 0.5, except the mass limit $M>10^{14}M_{\odot}h$. However, for the NN-PDF, the skewness is between $-0.2$ to $0.1$. Accordingly, we conclude that it is plausible to fit the SC-PDF with the skew-normal distribution. \\
In Fig.~\ref{fig:s} and Fig.~\ref{fig:l} we represent the mean and variance of the distributions. In Fig.~\ref{fig:s}, we plot the mean of the SC-PDF, $s_1$, with respect to a) $s_2$ and b) mass limit ($\log(M/M_{\odot})$). In Fig.~\ref{fig:l}, we show the logarithmic mean of the NN-PDF, $l_1$, with respect to a) $l_2$ and b) mass limit ($\log(M/M_{\odot}h)$). In both figures, the blue circles represent redshift zero, the orange squares correspond to redshift 0.5, and the green triangles show redshift 1. For the sake of clarity in presentation, in panel (b) of the figures, the mass scales are shifted slightly with respect to each other for the redshifts samples. There is one to one correspondence between the panels (a) and  (b).  In panel (a),  the mass limits can be derived using panel (b).  In Fig.~\ref{fig:s}, we study the different sample's similarities or differences. The SC-PDF $s_1$ and $s_2$ are similar for the different redshift samples with a fixed mass scale. Note that $s_2$  reveals the difference better than $s_1$. 
For the NN-PDF, based on Fig.~\ref{fig:l}, for the low mass scales, $l_1$ versus $l_2$ is almost the same, while for larger masses, we have more scatters. $l_2$ is a better probe to distinguish the samples, while $l_1$ is almost constant over a specific mass-scale. Note that in the corresponding figures, higher mass scales have larger error bars which is due to small number of massive halos in N-body simulation data-sets. \\
In Fig.~\ref{fig:s} panel (a) there is a positive correlation between $s_1$ and $s_2$ which can be approximated by a linear relation. Accordingly, for each specific redshift, a linear function can be fitted as a proxy for comparing different cosmological models. 
The relation between $l_1$ and $l_2$  is more complicated for the NN-PDF which is shown in Fig.~\ref{fig:l} panel (a). For the lower mass scales $\{11,11.5\}$,  $l_1$ and $l_2$ are anti correlated, while for the higher mass scales $\{12.5,13,13.5\}$ they become correlated. The mass scale $10^{12}$ is the turning point according to this plot.\\
{{ As we mentioned before the VP and the SC-CDF/PDF, are complimentary and known functions in the literature. Several studies has been done during past years to calculate the VP function in both simulations and observational data-sets. However, these studies often focus on calculating $\ln(P_0)/nV$ to observe universality of the VP function. As a specific example, the SC-CDF in the Fig.~\ref{fig:CDF} is compatible to the peaked-CDF in Fig.4 of \cite{Banerjee:2020umh}. In this direction, in our work we calculate the SC-PDF/CDF. Also, we should note that we study the less studied NN-PDF/CDF and J-function in this work. \cite{Kerscher:1999gt} have calculated the J-function for Matern cluster process, galaxies and mock samples. In this section, we show the updated results for the SC-CDF/PDF, NN-CDF/PDF, and J-function in terms of the dimensionless variable $x$, using high resolution N-body simulation data. We also point out three conclusive new statements:\\
1) {{Similarity argument}}: We find that for a specific mass scale, the mean of the SC-CDF is almost independent of redshift till $z=1$. \\
2) {{Distribution argument}}: Exploring the general behaviour of distributions, the SC-PDF is nearly skew-normal, while the NN-PDF is nearly log-normal.  \\
3) {{Moments dependency}}: There is an approximately linear relation between the mean and variance, however for the NN-CDF, we have a bi-modal behaviour in the moments.}}\\
In the next section, we continue our discussion using the observational data.

\section{Observations} \label{Sec4}
In this section, we study the SC-CDF/PDF, NN-CDF/PDF, and J-function for the galaxy group catalogues. Here, we are going to find the capability of the new probes to study clustering of galaxy groups.
Galaxy groups are the luminous tracers of large mass dark matter halos. Accordingly, we can compare the results with  N-body simulations. However, the exact comparisons need a more sophisticated approach to consider all the complication raised from the redshift space distortion, baryonic effects, and dark matter-luminous matter bias. {{As a complementary step toward studying this complex comparison, we study mock galaxy catalogues as well.}} 
\subsection{Sample selection: Galaxy groups and Mock catalogues}
\begin{table}
\centering
\caption{Different volume-limited catalogues are represented. The conditions on magnitude and redshift are presented in column one and two, respectively. $r_{\star} = r/(nV)^{1/3}$ is shown in column three. We report the number of groups in column four.}
\label{Ta: Galaxy groups}
\begin{tabular}{cccc}
\hline
$M_{r,\text{lim}}$ & $z_{\text{lim}}$ & $r_{\star}$ (Mpc/h) & N \\
\hline
\hline
-18 & 0.045 & 3.891 & 5258\\
-18.5 & 0.057 & 4.256 & 7761\\
-19 & 0.071 & 4.698 & 11352\\
-19.5 & 0.089 & 5.116 & 16517\\
-20 & 0.110 & 6.225 & 17301\\
-20.5 & 0.136 & 7.788 & 16444 \\
\hline
\end{tabular}
\end{table}
\begin{table}
\centering
\caption{{{The details of central galaxy samples from mock galaxy catalogue. The conditions on magnitude is written in column one. $r_{\star} = r/(nV)^{1/3}$ is shown in column two.}}}
\label{Ta: Mock}
\begin{tabular}{cc}
\hline
$M_{r,\text{lim}}$ & $r_{\star}$ (Mpc/h) \\
\hline
\hline
-19  & 2.679\\ 
-19.5 & 3.041\\ 
-20  & 3.691\\ 
-20.5 & 4.630\\ 
\hline
\end{tabular}
\end{table}
In this work, we use the galaxy group catalogues of \cite{tempel2014flux}. They provide flux-limited and volume-limited ones from SDSS-DR10 (\cite{york2000sloan,Ahn:2013gms}) data, and the upper limit on redshift is considered z=0.2. \cite{tempel2014flux} used a modified version of FoF to find the galaxy groups, in which the linking-length is redshift dependent. The linking-length is defined as the average over the NN-PDF of the galaxies in a sample. The idea behind this choice is to make sure that the galaxies find the nearest neighbour in their group, and isolated galaxies that are in the tail of the distribution, do not change the mean of the NN-PDF (and linking length) significantly. Accordingly, the average over the NN-PDF is a reasonable estimation for the distances between galaxies in a group. It is interesting to note that, for the volume-limited catalogues, the linking-length is approximately constant over redshift.\\ 
In this work, we use six volume-limited catalogues from \cite{tempel2014flux}. The details of the catalogues are shown in Table~(\ref{Ta: DM samples}). In the first column, we show the limit on the magnitude $M_{r,\lim}$, in the column two, we report the limit on redshift, and the column three is $r_{\star}$ for each catalogue in unit of $\text{Mpc/h}$.\\
In order to find the physical distances to the galaxy groups we need  to assume a background cosmology. We set the cosmological parameters similar to Section~\ref{Sec3}, in agreement with Planck \citep{Ade:2013zuv}.\\
\begin{figure}
\centering
\includegraphics[scale=0.30]{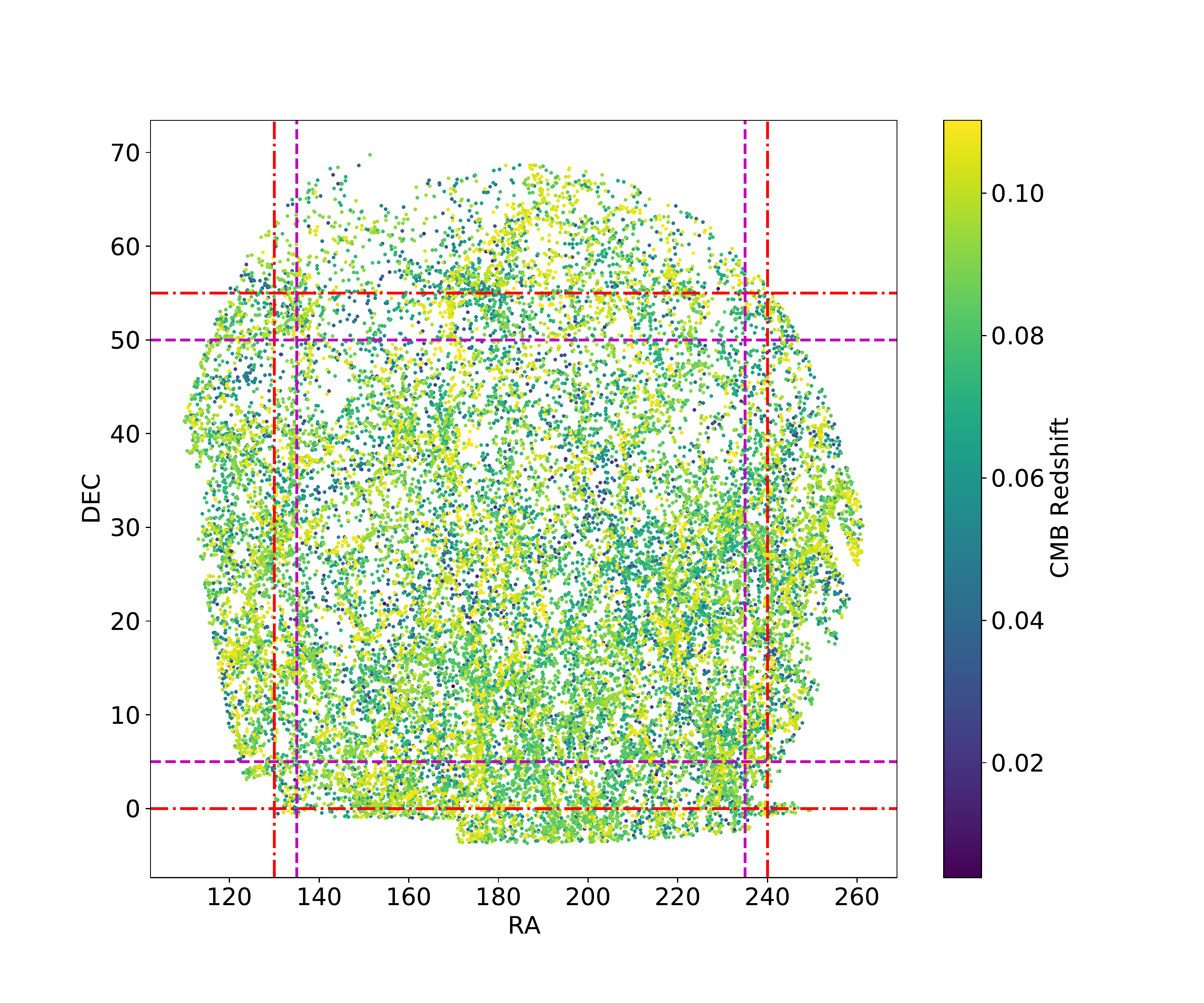}
\caption{{{The spatial position of galaxies used in this work. We only consider galaxies which are inside the square-frame indicated with dot-dashed red lines.}}}
\label{fig:SDSS}
\end{figure}
To avoid boundary problems in sample selection, we use an area of $0<Dec<55$, $130<Ra<240$. {{In Fig.~\ref{fig:SDSS},  the spatial position of all galaxies is plotted. The redshift of each galaxy is represented with colour. The square area which is enclosed with red dot-dashed lines, is the area being used in this work. We also use the inner square, which is represented with dashed purple lines, and is 5 degrees smaller in each direction, with respect to the main square. This is done to examine the effects of the boundary condition, after selection. However no significant change is detected for the inner area, so we only report the results of the main one.}}
In Table~(\ref{Ta: DM samples}), column four, we report the number of groups for each volume-limited catalogues, after selection. To estimate the errors, we use a conservative approach similar to Section~\ref{Sec3} and divide the sample into four equivalent sub-samples and calculate the mean and standard deviation of each sub-samples.\\
{{Mock catalogues are useful tools to study observational data \citep{Mao2018,Behroozi2019,Alam2021,Zhao2021}.
In our work, we use the mock galaxy catalogue from \cite{Paranjape:2021zia}. They find halos of an N-body simulation with ROCKSTAR halo-finder \citep{behroozi2012rockstar}, and populate halos with the mock central and satellite galaxies. The population of galaxies is based on an updated halo occupation distribution model. The host halo of each central galaxy is also baryonified. For our work, we select only the mock central galaxies, because they are positioned on the centre-of-mass of host halos. This helps to interpret results of the mock catalogue as a middle stage between halos and galaxy groups. We make four samples by considering limits on magnitude of the mock central galaxies. Details of these samples are shown in Table~(\ref{Ta: Mock}). Column one is the limit on the magnitude while column two represents $r_{\star}$.
The details of the original N-body simulation is as follows\fhn{:} a periodic cubic box simulation at redshift zero, which has $300$ Mpc/h length and $1024^3$ particles. The cosmological parameters of the simulation is set to WMAP7 best-fit (Table 1 of \cite{Paranjape:2020wuc}). The method for finding the NN-CDF, SC-CDF and errors are similar to Section~\ref{Sec3}.}} 
\subsection{Results}
\begin{figure}
\centering
\includegraphics[scale=0.3]{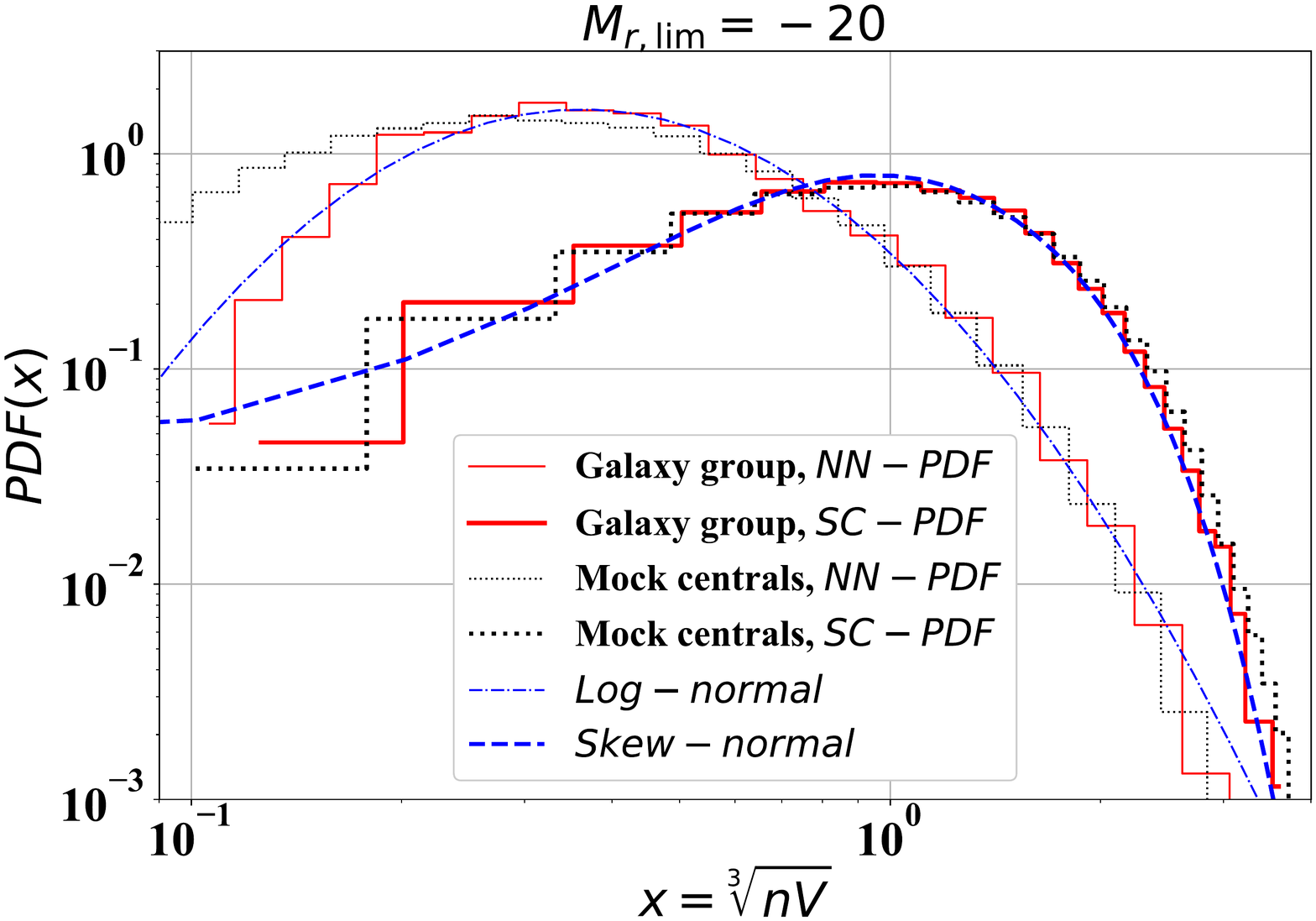}
\caption{The NN-PDF and the SC-PDF for galaxy groups are plotted versus the dimensionless variable $x$ for a volume limited catalogue. The condition on magnitude is $M_{r,\text{lim}}=-20$. Thin red line refers to NN-PDF and the thick red line refers to SC-PDF. Dot-dashed thin blue line is the theoretical log-normal curve equation~(\ref{Eq: lognormal}) with $\mu =-0.68,\sigma=0.58$ and dashed thick blue line is skew-normal curve equation~(\ref{Eq: Skewnormal}) with $\mu =0.56,\sigma=0.8,\alpha=2.6$. {{The thin black step dotted and the thick black step dotted plots represent the NN-PDF and SC-PDF of central galaxy samples of the mock galaxy catalogue, respectively. The condition on magnitude of the central galaxies is as $M_{r,\text{lim}}=-20$.}}}
\label{fig:HistG}
\end{figure}
In Fig.~\ref{fig:HistG}, we plot the SC-PDF (thick red line) and the NN-PDF (thin red line) for $M_{r,\text{lim}}=-20$ as a function of $x$, for the galaxy groups. The SC-PDF peak is around $1.2$, and the NN-PDF peak is at $0.5$. They correspond to physical comoving lengths of $7.5$ Mpc/h and $3.1$ Mpc/h, respectively. Comparing the galaxy groups with simulations is challenging because of the differences between statistics of DM halos and galaxy groups. However, the general behaviour of dark matter distributions in N-body simulations (see Section~\ref{Sec3}) can be used as a hint for galaxy distributions. For example, we showed in Fig.~\ref{fig:Hist} that the SC-PDF is nearly skew-normal, and the NN-PDF approximately follows a log-normal distribution for dark matter halos. This approximation seems viable for galaxy groups as well. In Fig.~\ref{fig:HistG} dashed blue thick line, shows skew-normal probability distribution function (equation~(\ref{Eq: Skewnormal})) with parameters $\mu =0.56,\sigma=0.8,\alpha=2.6$ and dot-dashed blue thin line is a log-normal probability distribution function (equation~(\ref{Eq: lognormal})) with parameters $\mu =-0.68,\sigma=0.58$. Here, similar to Section~\ref{Sec3}, we find the parameters when the distributions approximately follow the theoretical distributions. Similar to DM halos, the NN-PDF peaks in smaller scales compared to the SC-PDF. Again, we conclude that the NN-PDF probes smaller scales in comparison with the SC-PDF which probes larger scales.\\
{{In Fig.~\ref{fig:HistG} we show the results of the mock central galaxies with thin/thick black step dotted lines, which correspond to the NN-PDF/SC-PDF. The condition on the magnitude of the mock galaxies is $M_{r,\lim}=-20$. There is a good agreement between the statistics of the SC-PDF of the mock central galaxies and the galaxy groups. However, there is a difference between the two curves of NN-PDF for small distances $x<0.2$. Making mock galaxy group catalogues with both central and satellite galaxies could relax the difference. Also, considering the neglected baryonic effects and the redshift space distortion (RSD) could be a solution.}}\\
\begin{figure}
\centering
\includegraphics[width=0.43\textwidth]{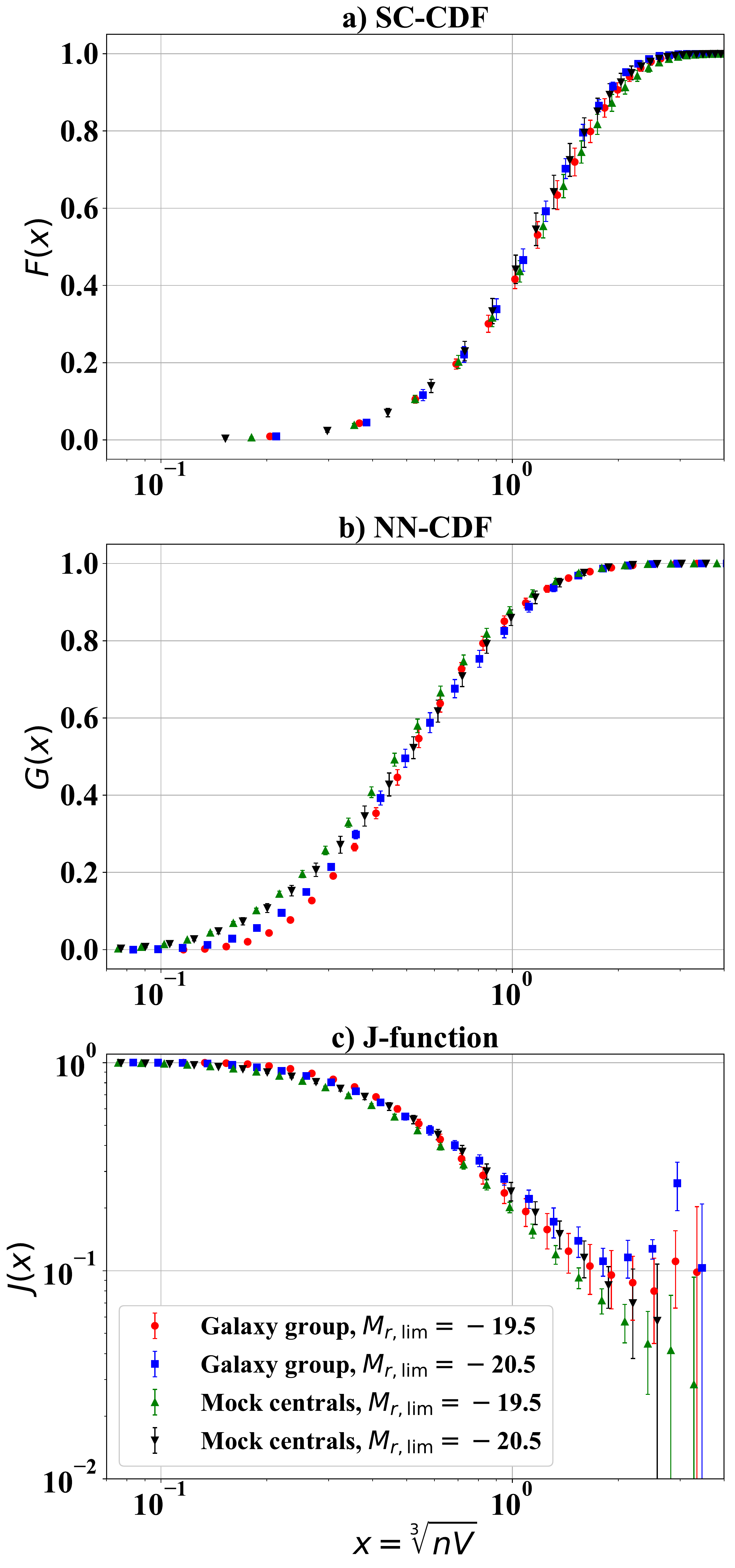}
\caption{In this figure, in panels a) $F(x)$, SC-CDF, b) $G(x)$, NN-CD and c) $J(x)=\dfrac{1-G(x)}{1-F(x)}$ versus $x=\sqrt[3]{nV}$ are plotted for the galaxy groups of the volume limited catalogue and the central galaxies of the mock galaxy catalogue. For the galaxy groups, red circles represent $M_{r,\text{lim}}=-19.5$ and blue squares represent $M_{r,\text{lim}}=-20.5$. For the central galaxies, green up-pointing triangles show $M_{r,\text{lim}}=-19.5$ and black down-pointing triangles show $M_{r,\text{lim}}=-20.5$ .}
\label{fig:CDFG}
\end{figure}
{{In Fig.~\ref{fig:CDFG} we plot the  SC-CDF in panel (a) and the NN-CDF in panel (b) for the galaxy groups and the mock central galaxies. In panel (c), we plot the J-function where the samples are selected as $M_{r,\text{lim}}=\{-19.5,-20.5\}$ (see Table~(\ref{Ta: Galaxy groups}) and Table~(\ref{Ta: Mock})). We represent galaxy groups with a red circle and blue square, while mock galaxies are shown with green up-pointing and black down-pointing triangles, (corresponding to different magnitude limit, respectively).}} The similarity of different samples within error bars can be deduced based on Fig.~\ref{fig:CDFG}. Note that we do not consider any limit on the mass of the galaxy groups.
So low mass and high mass groups are present in the volume-limited catalogues. This situation is comparable with the low mass dark matter halo samples of Section~\ref{Sec3}. In near future, the number of observed galaxy groups will increase. Accordingly, it would be possible to set additional conditions on the samples, such as dynamical mass limit. This can help us to break the degeneracy between different samples.\\
{{ By comparing the mock central galaxies with the galaxy groups, the results agree within error-bars, when we use the SC-CDF. However, the NN-CDF shows some deviation at small x. These results are consistent with Fig.~\ref{fig:HistG}.}}\\
\begin{table*}
\centering
\caption{We report the moments of the SC-PDF for the samples of Table~(\ref{Ta: Galaxy groups}) in the first six rows, and Table~(\ref{Ta: Mock}) in the last four rows (marked with *), calculated by equation~(\ref{Eq: moments}). The condition on the magnitude is presented in the column one. In the columns 3, 4, 5, 6 we report the quantities $s_1$, $s_2$, $\tilde{s}_3$ and $\tilde{s}_4$, respectively. }
\label{Ta: SCFG}
\begin{tabular}{ccccc}
\hline
$M_{r,\text{lim}}$ & $s_1$ & $s_2$ & $\tilde{s}_3$ & $\tilde{s}_4$ \\
\hline
\hline
18 & $1.202 \pm 0.045$ & $0.586 \pm 0.029$ & $0.648 \pm 0.070$ & $3.264 \pm 0.165$ \\
18.5 & $1.188 \pm 0.059$ & $0.564 \pm 0.030$ & $0.633 \pm 0.056$ & $3.324 \pm 0.179$ \\
19 & $1.157 \pm 0.031$ & $0.547 \pm 0.019$ & $0.679 \pm 0.066$ & $3.457 \pm 0.225$ \\
19.5 & $1.198 \pm 0.045$ & $0.558 \pm 0.024$ & $0.563 \pm 0.049$ & $3.120 \pm 0.171$ \\
20 & $1.177 \pm 0.037$ & $0.544 \pm 0.013$ & $0.620 \pm 0.044$ & $3.262 \pm 0.137$ \\
20.5 & $1.165 \pm 0.035$ & $0.519 \pm 0.010$ & $0.518 \pm 0.128$ & $3.159 \pm 0.512$ \\
\hline
*19.0 & $1.239 \pm 0.047$ & $0.608 \pm 0.03$ & $0.738 \pm 0.074$ & $3.531 \pm 0.261$ \\ 
*19.5 & $1.221 \pm 0.048$ & $0.592 \pm 0.031$ & $0.702 \pm 0.098$ & $3.434 \pm 0.35$ \\ 
*20.0 & $1.201 \pm 0.056$ & $0.57 \pm 0.034$ & $0.649 \pm 0.082$ & $3.279 \pm 0.267$ \\ 
*20.5 & $1.168 \pm 0.061$ & $0.545 \pm 0.036$ & $0.634 \pm 0.094$ & $3.323 \pm 0.28$ \\ 
\hline
\end{tabular}
\end{table*}
\begin{table*}
\centering
\caption{The logarithmic moments of the NN-PDF related to the samples of Table~(\ref{Ta: Galaxy groups}) in the first six rows, and Table~(\ref{Ta: Mock}) in the last four rows (marked with *), calculated by equation~(\ref{Eq: logmoments}). The condition on the magnitude is shown in column 1. In columns 3, 4, 5, 6 the quantities $l_1$, $l_2$, $\tilde{l}_3$ and $\tilde{l}_4$ are respectively represented. }
\label{Ta: NNDG}
\begin{tabular}{ccccc}
\hline
$M_{r,\text{lim}}$ & $l_1$ & $l_2$ & $\tilde{l}_3$ & $\tilde{l}_4$ \\
\hline
\hline
18 & $-0.672 \pm 0.034$ & $0.554 \pm 0.018$ & $0.242 \pm 0.043$ & $2.768 \pm 0.126$ \\
18.5 & $-0.658 \pm 0.027$ & $0.56 \pm 0.018$ & $0.216 \pm 0.042$ & $2.728 \pm 0.119$ \\
19 & $-0.658 \pm 0.031$ & $0.554 \pm 0.009$ & $0.157 \pm 0.083$ & $2.664 \pm 0.091$ \\
19.5 & $-0.663 \pm 0.024$ & $0.561 \pm 0.013$ & $0.151 \pm 0.048$ & $2.679 \pm 0.064$ \\
20 & $-0.679 \pm 0.024$ & $0.583 \pm 0.015$ & $0.107 \pm 0.03$ & $2.574 \pm 0.046$ \\
20.5 & $-0.679 \pm 0.027$ & $0.618 \pm 0.016$ & $-0.013 \pm 0.051$ & $2.527 \pm 0.047$ \\
\hline
*19.0 & $-0.803 \pm 0.027$ & $0.665 \pm 0.010$ & $-0.157 \pm 0.027$ & $2.696 \pm 0.035$ \\ 
*19.5 & $-0.789 \pm 0.026$ & $0.665 \pm 0.009$ & $-0.176 \pm 0.031$ & $2.692 \pm 0.039$ \\ 
*20.0 & $-0.768 \pm 0.033$ & $0.661 \pm 0.013$ & $-0.205 \pm 0.049$ & $2.689 \pm 0.066$ \\ 
*20.5 & $-0.731 \pm 0.044$ & $0.665 \pm 0.017$ & $-0.271 \pm 0.067$ & $2.736 \pm 0.100$ \\ 
\hline
\end{tabular}
\end{table*}
We calculate moments of the SC-PDF in Table~(\ref{Ta: SCFG}) and logarithmic moments of the NN-PDF in Table~(\ref{Ta: NNDG}). In the column one of the tables, we present magnitude limit. In the rest of the four columns $s_1$, $s_2$, $\tilde{s}_3$ and $\tilde{s}_4$ are presented for the SC-PDF $l_1$, $l_2$, $\tilde{l}_3$ and $\tilde{l}_4$ for the NN-PDF. The errors on the moments are less than $5\%$ for the mean and variance. {{In both tables, first six rows are related to the galaxy groups and the other four rows are for the mock galaxy centrals, which are marked with *.}}\\
The SC-PDF skewness is around 0.5, and the NN-PDF has approximately 0.15 logarithmic skewness. (See the column four of the Tables~(\ref{Ta: SCFG},\ref{Ta: NNDG})). Similar to the halos, the NN-PDF skewness is smaller than the SC-PDF and kurtosis of the SC-PDF is approximately between 3.2 and 3.4. On the other hand, logarithmic kurtosis of the NN-PDF is approximately between 2.5 and 2.7. This shows that the SC-PDF can be considered as a nearly normal distribution while the NN-PDF is nearly log-normal, which is consistent with Fig.~\ref{fig:HistG} and the SMDPL simulation.\\
{{All the samples of the mock central galaxies and the galaxy groups have equal moments of the SC-PDF, considering the error-bars, except the one with $M_{r,\lim}=-19$. But this is not the case for the NN-PDF, which is consistent with our aforementioned results.}}\\
\begin{figure}
\includegraphics[scale=0.3]{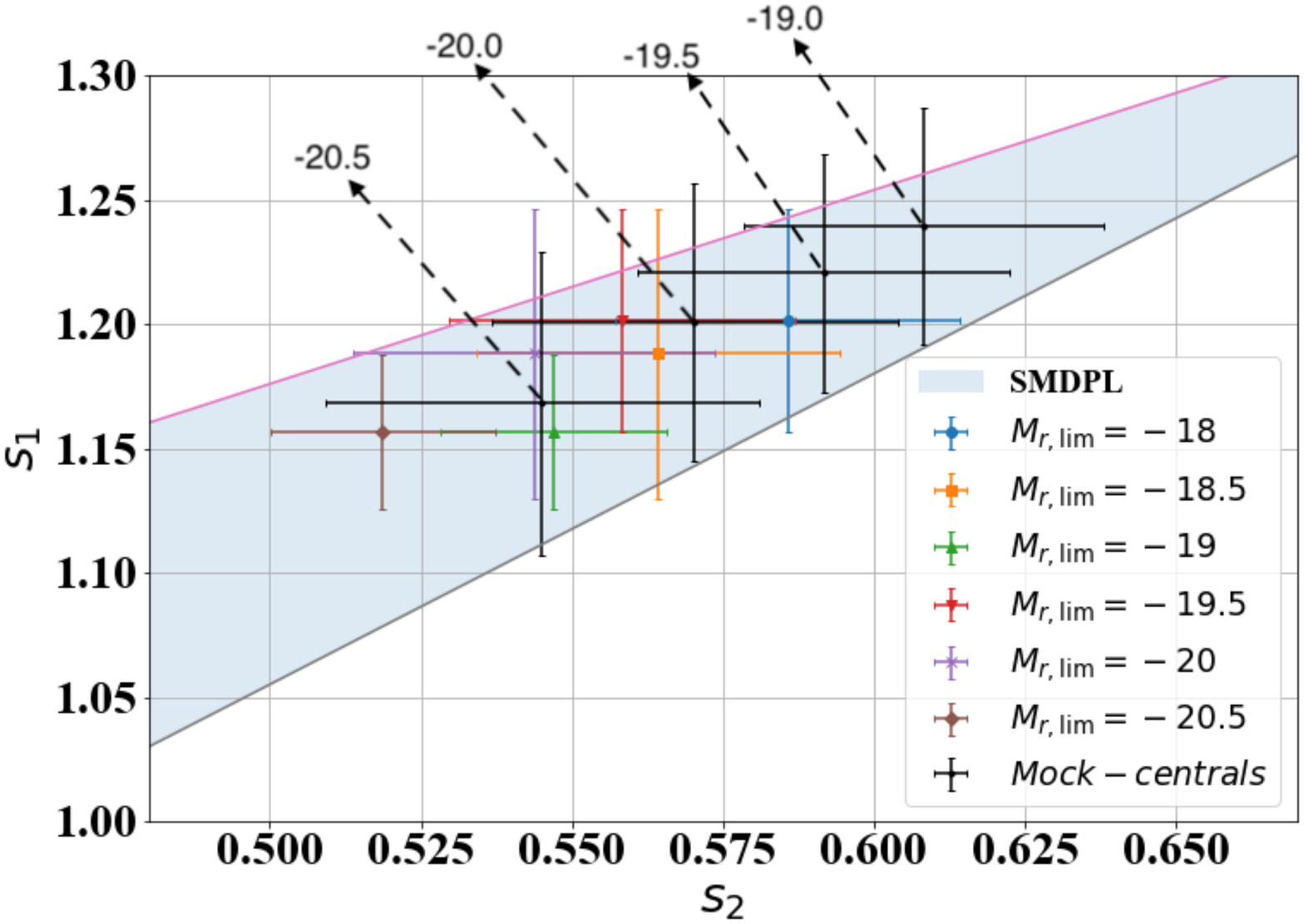}
\caption{The mean of the SC-PDF for galaxy groups of the volume limited catalogue, $s_1$, is plotted for different magnitude limits with respect to $s_2$.
We show sample $M_{r,\text{lim}}=-18$ with the blue circle, $M_{r,\text{lim}}=-18.5$ with the orange square, $M_{r,\text{lim}}=-19$ with the green up-pointing triangle, $M_{r,\text{lim}}=-19.5$ with the red down-pointing triangle, $M_{r,\text{lim}}=-20$ with the purple and $M_{r,\text{lim}}=-20.5$ with the brown diamond. The result of $\Lambda$CDM halos from the SMDPL simulation is shown by soft blue region. {{We also represent results for the central galaxies of the mock galaxy catalogues with black dots. The conditions on the magnitude of central galaxies are shown inside the figure.}}} \label{fig:sG}
\end{figure}
In Fig.~\ref{fig:sG} we plot $s_1$ with respect to $s_2$ for the ten samples of Table~(\ref{Ta: SCFG}). For galaxy groups, blue circle shows $M_{r,\text{lim}}=-18$, orange square shows $M_{r,\text{lim}}=-18.5$, green up-pointing triangle shows $M_{r,\text{lim}}=-19$, red down-pointing triangle shows $M_{r,\text{lim}}=-19.5$, purple $x$-symbol shows $M_{r,\text{lim}}=-20$ and brown diamond shows $M_{r,\text{lim}}=-20.5$. Soft blue region is extracted from the SMDPL which we plot it in Fig.~\ref{fig:s}. The moments derived from galaxy groups shows consistency with the soft blue region from the SMDPL. This result is compelling, considering the difference between samples of the dark matter halos of SMDPL simulation and the galaxy groups catalogue.
The volume-limited galaxy groups have the condition on their magnitudes, while the SMDPL halo samples have constrained on their masses.\\
{{In Fig.~\ref{fig:sG}, four black dots are related to the mock central galaxies with different conditions on their magnitude limit (which are presented inside the figure). It is clear that the result for the mock galaxies and halos are consistent. Moreover, we can see the linear relation between the mean and variance of the SC-PDF, with respect to the magnitude limit of the mock central galaxies.}} \\
The uncertainties are substantial, both for simulation and observational data-sets. However, in the light of precise future galaxy surveys \citep{Amendola:2016saw} and technological advances in numerical simulations \citep{Adamek:2016zes, Hassani:2019lmy}, one could reduce the errors significantly. Thus a more precise comparison between the data and simulation will be available in near future.\\
\section{Conclusions and Remarks} \label{Sec5}
The standard model of cosmology is well tested in the last two decades in linear regime mainly, using two-point statistics. With the upcoming surveys we are going to observe the Universe in deeply non-linear regime. Accordingly, the main challenge would be to compare the theory with observation in non-linear scales. In this work, we mainly focus on the idea of using complimentary statistical probes to study  non-linear scales.
In non-linear scales, the matter distribution is far from Gaussian, and we can approximate it by a log-normal distribution. At these scales, study of the statistics that depend on higher-order correlation functions is experimentally and theoretically motivated. At first glance, they can improve constraints on the cosmological parameters for a given data. Moreover, the two other advantages of these functions are 1) Finding the universality behaviours. For example, the scaling relation for the void function gives information about the statistics of matter density and the form of $n$-point correlation function of halos/galaxies. 2) Finding differences between samples which leads us to extract more information in order to constrain the parameters and to distinguish cosmological models. For an example, breaking the degeneracy between bias and $\sigma_8$ parameters \citep{Banerjee:2020umh}.\\
In this work, we used the SMDPL simulation from the MultiDark project. Also, we use the galaxy groups from the volume-limited Tempel catalogue, {{as well as the mock central galaxies from a mock galaxy catalogue}}. We calculated the Spherical Contact (SC) distribution function, Nearest Neighbour(NN) distribution function, and the J-function in simulations and observational data. We choose the dark matter halo samples from the SMDPL by introducing limits on mass and redshift. We use the galaxy group sample {{and the mock central galaxies}} with limits on the magnitude. \\
Our results suggest that, the SC-CDF/PDF probes larger scales compared to the NN-CDF/PDF. Accordingly, the  NN-CDF reveals difference between the samples. 
For specific mass scale, different redshifts have similar $s_1$, $l_1$ but they have different $s_2$, $l_2$ for the SC-PDF and NN-PDF respectively. We also obtain similar results for galaxy groups.
Our main findings are summarised as follows:\\
\begin{itemize}
\item In both the SMDPL and galaxy groups, the SC-PDF distribution is nearly skew-normal and the NN-PDF is nearly log-normal. The NN-PDF is more sensitive to non-linear clustering than the SC-PDF. The non-unity of the J-function comes from the difference between the NN-PDF and SC-PDF.
\item For a specific mass scale and different redshifts, the mean of distribution is the same for both the SC-PDF and NN-PDF. As a result, we can use the variance to distinguish between samples. However, the NN-CDF reveals the differences more clearly.
\item In the SMDPL, for the SC-PDF, a linear relation between $s_1$ and $s_2$ is found. The results for the galaxy groups are also consistent with the SMDPL considering the error bars. Using a more precise approach of error finding and considering the simulations with larger box sizes, errors will be reduced significantly and we can compare the $s_1$, $s_2$ relation between the simulations and data better. The relation between $l_1$ and $l_2$ is complicated and the trend changes in mass scales around $\sim 10^{12}M_{\odot}h$.
\item We find the mean and variance of the SC-PDF/NN-PDF for galaxy groups within 5 percent errors.
\item Some samples have similar SC-CDF but have different NN-CDF. This suggests a complementary role of the SC-CDF and NN-CDF and show that they contain different information. 
\item The J-function is equal to the first conditional correlation. It contains information about clustering. We calculated the J-function for all the samples, and showed that it is sensitive  to the mass scale. In the larger mass scales, we have less clustering as their J-function is closer to one. 

\item {{We find a good agreement between the results for mock central galaxy samples and galaxy groups, when we focus on the SC-PDF (at least for $M_{r,\lim}=\{-19.5,-20,-20.5\}$). However, for the case of the NN-PDF there is a difference between samples at small distances $x<0.2$. Making mock galaxy group catalogues or considering some ignored baryonic effects and redshift space distortions may relax this difference.}}
\end{itemize}
For future remarks, note that we usually test cosmological models to find the best fit of the parameters. This is not easily done at non-linear scales. However, our results suggest a possible intrinsic length scale in the form of the mean of the SC-PDF. Based on our findings in this work, the first moments of the SC-PDF is redshift independent. The universality of the SC-CDF and the redshift independency could be a promising result to define a standard length scale for non-linear regime.
As a future work, one can study the universality of the J-function, which is a result of the scale-invariant correlation function.
Finding the standard lengths from non-linear structures is important, as they can be used as a ruler. These standard lengths potentially can reveal tensions between distinct data sets (for example the $H_0$ tension in \cite{Riess_2019}). 
We also can search for cosmological models which break the universality or similarity.
\section*{Acknowledgements}
We are grateful to Ali Akbar Abolhassani, Julian Adamek, Jean-Pierre Eckmann, Nima Khosravi, Martin Kunz, David Mota, Aseem Paranjape, Sohrab Rahvar, Saeed Tavasoli, and Cora Uhlemann for many fruitful discussions and insightful comments on the manuscript. We thank Aseem Paranjape, for providing the mock galaxy catalogue. We also thank the anonymous referee for his /her insightful comments.
SB is partially supported by Abdus Salam International Center of Theoretical Physics (ICTP) under the junior associateship scheme. This research is supported by Sharif University of Technology Office of Vice President for Research under Grant No. G960202.\\
The CosmoSim database used in this paper is a service by the Leibniz-Institute for Astrophysics Potsdam (AIP).
The MultiDark database was developed in cooperation with the Spanish MultiDark Consolider Project CSD2009-00064\\
The computations were performed at University of Geneva on the Baobab and Yggdrasil clusters. Part of the computations for this paper were performed on the Euclid Cluster at the Institute of Theoretical Astrophysics at University of Oslo. 
\section*{Data Availability}
In this paper, we used data from the Small Multidark Planck simulation\footnote{doi:10.17876/cosmosim/smdpl/}. The FoF halo catalogues are available at the CosmoSim database (\href{https://www.cosmosim.org/cms/simulations/smdpl/}{www.cosmosim.org/cms/simulations/smdpl/}). For the observational section, we used the Tempel 2014 galaxy catalogues from the CosmoDB database (\href{http://cosmodb.to.ee/cms/documentation/sdss_dr10/}{www.cosmodb.to.ee/cms/documentation/sdss\_dr10/}).
\\
{{The mock catalogue data is received from the authors of the paper  \cite{Paranjape:2021zia}. The codes developed in this work are available under request}}.



\bibliographystyle{mnras}
\bibliography{Bib} 




\appendix

\section{Universality of the Void Probability function} \label{App_Uni}
The relation between the SC-CDF and the NN-CDF to cosmological parameters and initial conditions is complicated. In this direction, universal features of non-linear matter density help us extract information from deeply non-linear scales. For example, \cite{white1979hierarchy} proposed a scaling relation for the void probability (VP) function. In this section, we review the universality of the VP function. It results in a universal form for $\ln P_0/nV$. We also find a similarity for the mean of the SC-PDF in different redshifts.\\
We follow \cite{balian1989scale} and show the scaling relations from equation~(\ref{Eq: VP}), while assuming a scale invariant $n$-point correlation function as,
\begin{equation} \label{Eq: scale}
\xi_N(\lambda \textbf{r}_1,...,\lambda \textbf{r}_N) = \lambda^{-(N-1) \gamma } \xi(\textbf{r}_1,...,\textbf{r}_N),
\end{equation}
where $\gamma$ is a constant. This form of $n$-point correlation function has a theoretical and observational motivation\citep{Bernardeau:2001qr}. The scale invariant assumption, as well as the transitional and rotational symmetry leads to the following power-law relation for the two-point correlation function
\begin{equation}
r^\gamma \xi_2(r) = \text{constant.}
\end{equation}
Averaging over volume $V$ leads to a similar equation for the averaged two-point correlation function
\begin{equation} \label{Eq: AveCorr}
V^{\gamma/3} \int \dfrac{1}{V^2} \xi_2(\textbf{r}_1,\textbf{r}_2) \, dV_1 dV_2 = V^{\gamma/3} \bar{\xi}_2 = \text{constant}.
\end{equation}
It is straightforward to show that the scale invariance assumption results in a scaling relation for the conditional correlation function $\Xi_i$, equation~(\ref{Eq: conditional corr})
\begin{equation} \label{Eq: Uni}
\Xi_i(\textbf{r}_1,...,\textbf{r}_i;V) = \sum_j \dfrac{(-nV)^j}{j!} (\bar{\xi}_2)^{i+j-1} S_{i,j}(\textbf{r}_1,...,\textbf{r}_i),
\end{equation}
in which $S_{i,j}$ is
\begin{equation} \label{Eq: S}
S_{i,j}(\textbf{r}_1,...,\textbf{r}_i) =\int \xi_{i+j}((\bar{\xi}_2)^{1/\gamma} \textbf{r}_1,...,(\bar{\xi}_2)^{1/\gamma} \textbf{r}_{i+j}) \, \dfrac{dV_{i+1}}{V} ... \dfrac{dV_{i+j}}{V}.
\end{equation}
Then we show that $S_{i,j}$ is independent of volume and $\bar{\xi}_2$. In this case, the quantity $(nV)^{i-1}\Xi_i$ is a universal function of $nV\bar{\xi}_2$ and so different samples exhibit similar behaviour with respect to $nV\bar{\xi}_2$. Finally, the observational quantities e.g., $\ln P_0/(nV)$, which is equal to $\Xi_0/(nV)$ (see equation~(\ref{Eq: VP})), is treated as a universal function of $nV\bar{\xi}_2$. However, the universality happens only if $S_{i,j}$ is independent of volume $V$, the averaged two point correlation function $\bar{\xi}_2$ and also $\textbf{r}_i$ for $i>0$. This argument is intuitively grasped when we apply a changing variables as $d^3\rho_j = \dfrac{d V_{j}}{V}$. By this definition, in equation~(\ref{Eq: S}) the integrals are taken over spheres with radius of unity length and we have
\begin{equation}
S_{i,j} (\rho_1,...,\rho_i) =\int \xi_{i+j}( c \rho_1,...,c \rho_{i+j}) \, d^3\rho_{i+1} ... d^3\rho_{i+j},
\end{equation}
where $c=(\bar{\xi}_2)^{1/\gamma} V^{1/3}$ is a constant using equation~(\ref{Eq: AveCorr}). \cite{balian1989scale} examined the universality of $S_{0,j}$. For $S_{i,j}$ the arguments might not hold anymore. Finding an exact solution for $S_{i,j}$ is out of the scope of this work.\\
\citep{Croton:2004ac} investigated the universality of $\ln P_0/(nV)$, and they proposed negative binomial function as a good approximation for this function in
\begin{equation}
\ln P_0/(nV) = \dfrac{1}{nV\bar{\xi}_2} (1+\ln (nV\bar{\xi}_2)).
\end{equation}
Employing equation~(\ref{Eq: J}), we see that the J-function is equal to $\Xi_1$. It suggests that J-function is also a universal function. \\
In the following, we focus on variable $nV \bar{\xi}_2$. $\bar{\xi}_2$ calculated by equation
\begin{equation}
\bar{\xi}_2 = \dfrac{\langle N^2 \rangle -\langle N \rangle^2-\langle N \rangle}{\langle N \rangle^2},
\end{equation}
where $\sqrt{\langle N \rangle^2-\langle N \rangle^2}$ is the standard deviation of number count probability of the volume $V$. $\langle N \rangle$ is the mean number of particles found in volume $V$, which is equal to $nV$. In literature $\bar{N} \equiv nV$ is defined instead of $r$ variable \citep{Fry:2013pma}. However, we mainly focus on the length scales. We redefine $x^3 \equiv (\dfrac{4\pi}{3} n) r^3$ in which $V = 4\pi/3 r^3$ and throughout this work we investigate $x$ (dimensionless variable) dependency of the SC-CDF, the NN-CDF, and  the J-function. Finally, we mention that we find similarity for the VP  function versus dimensionless variable $x$ and not the well-known parameter $x\bar{\xi}_2$. This argument suggests a new form of universality in redshifts, which we will investigate in future works with more details.
\section{Error estimation} \label{APP_Sim}
In this appendix, we discuss the results of the other simulations of the "MultiDark" project \citep{riebe2013multidark}. Especially the MDPL and MDPL2. They are similar to SMDPL but with a larger box size of 1 Gpc/h. The MDPL and MDPL2 are different in their seed numbers. So in this section, we discuss the effect of the box size and the initial conditions. {{In the following we discuss the errors induced by the boundary condition.}} Here, similar to Section~\ref{Sec3}, we estimate the errors conservatively, where we divide the main box into 27 sub-boxes. We also present our approach to the error estimation in this appendix and with more details. \\
\begin{figure}
\centering
\includegraphics[width=0.43\textwidth]{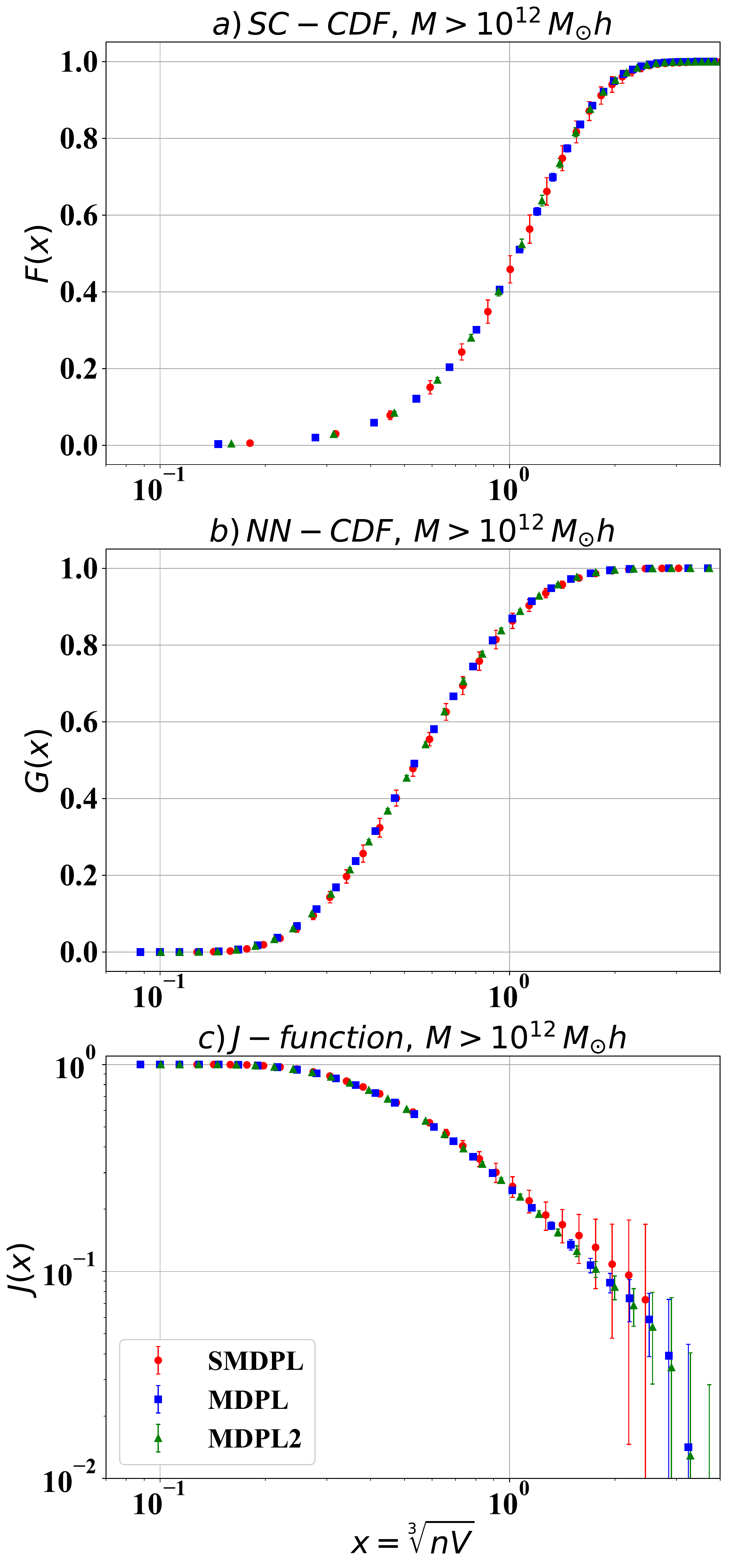}
\caption{We plot in panel (a) $F(x)$, SC-CDF, (b) $G(x)$, NN-CDF and (c) $J(x)=\dfrac{1-G(x)}{1-F(x)}$  with respect to $x=\sqrt[3]{nV}$ for DM halos from simulation. Mass limit is $10^{12}M_{\odot}h$ and $z=0$. Red circles represent the SMDPL simulation, blue squares correspond to MDPL and green triangles are for MDPL2.}
\label{fig:MDPL}
\end{figure}
In Fig.~\ref{fig:MDPL} we plot in panels (a) $F(x)$, (b) $G(x)$, and (c) $J(x)=\dfrac{1-G(x)}{1-F(x)}$ as a function of $x=\sqrt[3]{nV}$. In this figure, the mass limit is $M>10^{12} M_{\odot}h$ and $z=0$. Since the box size of the MDPL and MDPL2 simulations are larger, the error bars are smaller and barely recognisable in the figures. The red circles are for the SMDPL, for MDPL we use blue square, and the green triangles are for the MDPL2. 
Figures show good agreement between the same samples in different simulations.  \\
\begin{figure*}
\includegraphics[ width=1\textwidth]{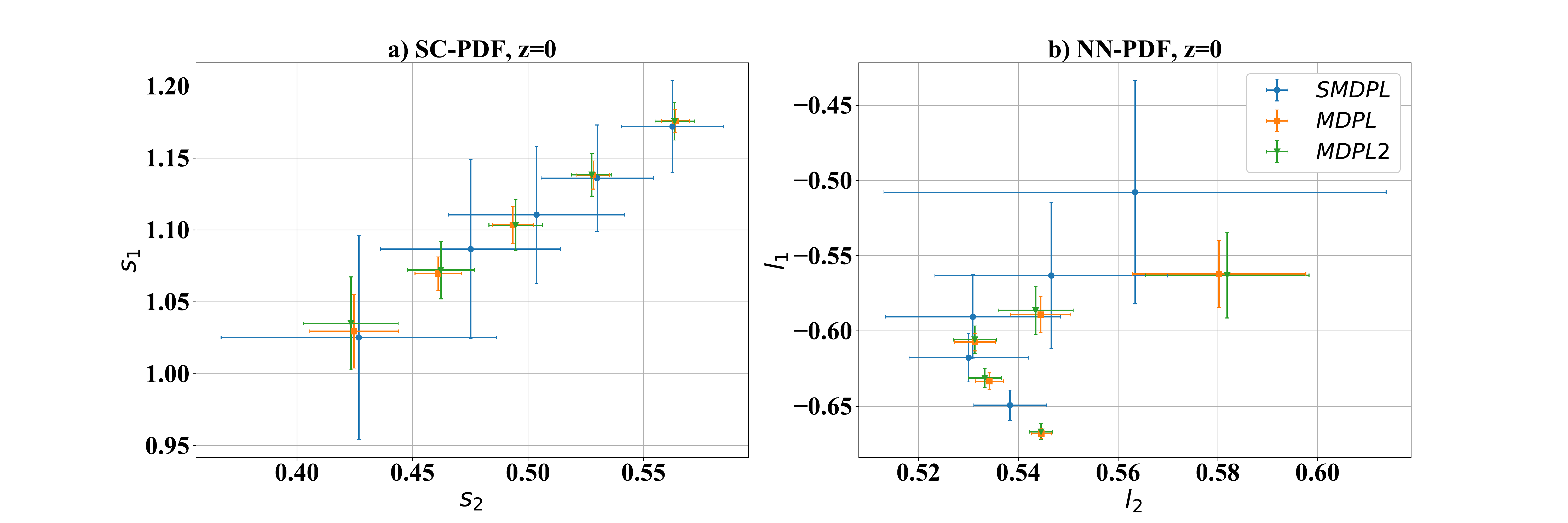}
\caption{The panel (a) $s_1$ with respect to $s_2$ and (b) $l_1$ with respect to $l_2$ is plotted for dark matter halos from different simulations. The blue circle is for the SMDPL, orange square is for MDPL, green triangle is for MDPL2. Points represent the mass limits $\log(M_{\lim}/M_{\odot}h)=\{12,12.5,13,13.5,14\}$. }
\label{fig:MDPL-moments}
\end{figure*}
In Fig.~\ref{fig:MDPL-moments} we plot in panel (a) $s_1$ as a function of $s_2$, for the SC-PDF and in panel (b) $l_1$ as a function of $l_2$, for the NN-PDF. In this figure, we only consider the case $z=0$, and the different points are related to different mass scales, $\log(M_{\lim}/M_{\odot}h)=\{12,12.5,13,13.5,14\}$. 
In Fig.~\ref{fig:MDPL-moments} moments of the SC-PDF is similar in the different simulations, which were predictable as the functions are independent of the box size. It is also true for moments of the NN-PDF. Except to the mass scale above $10^{12}M_{\odot}h$. For this mass scale $s_1$ is $-0.649 \pm 0.010$ for the SMDPL, $-0.668 \pm 0.003$ for the MDPL and $-0.666 \pm 0.005$ for the MDPL2. MDPL and MDPL2 agree. However, they are far from SDMPL results within one sigma confidence level. An explanation is systematic from boundary effects \cite{chiu2013stochastic}. In these scales, the statistical errors are small and comparable to the systematic errors. In the MDPL and the MDPL2, the number of halos is large, and the errors are negligible.\\
\begin{figure}
\centering
\includegraphics[scale=0.3]{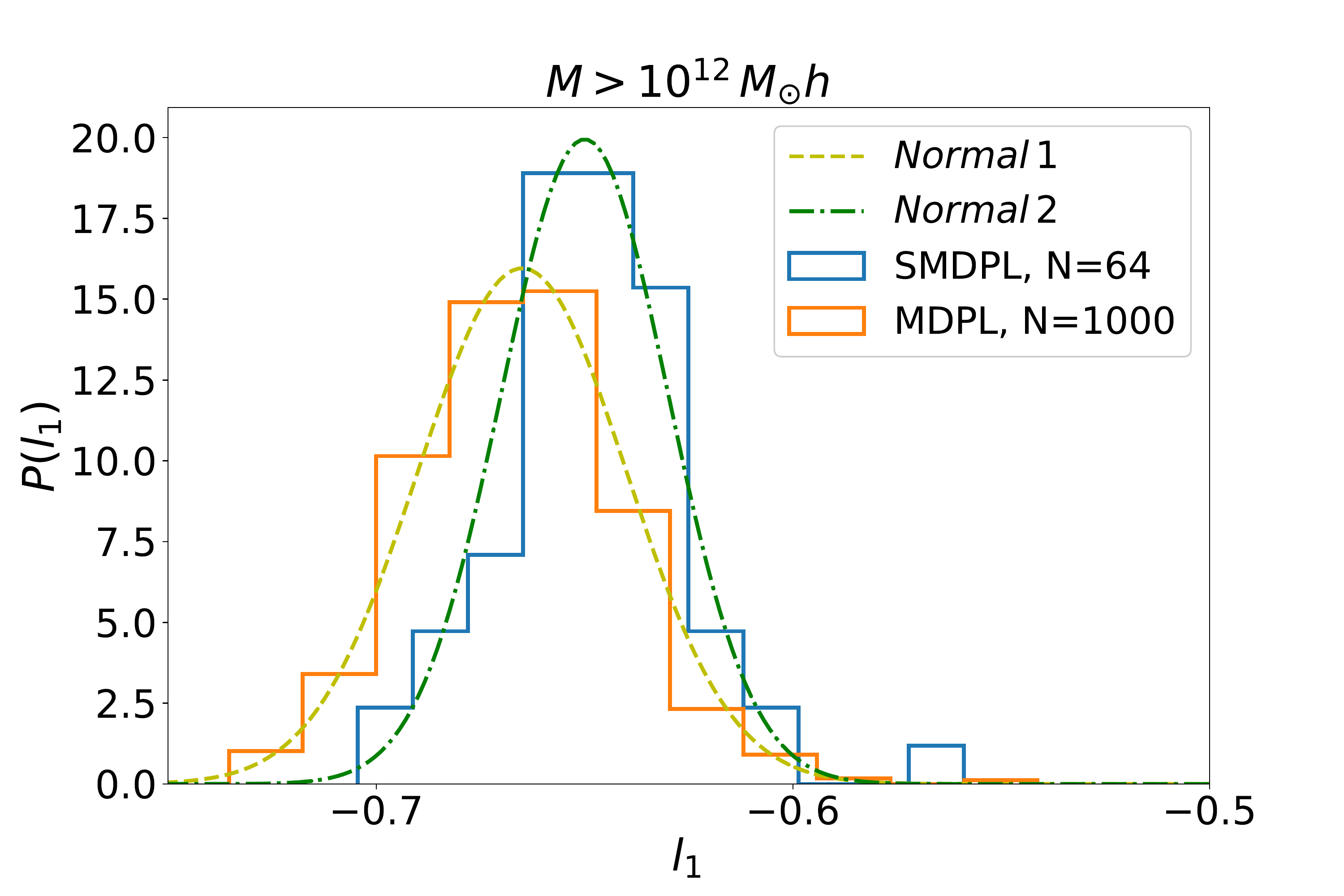}
\caption{The NN-PDF logarithmic mean, $l_1$ is plotted for different sub-boxes. The SMDPL simulation with 64 sub-boxes is plotted in blue, and the MDPL simulation with 1000 sub-boxes in a solid orange line. The estimated normal distributions are plotted with dot-dashed green and dashed yellow lines, respectively.}
\label{fig:error}
\end{figure}
To find the statistical errors, we divided each simulation box into different sub-boxes and reported the variance as an error. Accordingly, we divide the MDPL to $10^3$ and the SMDPL to $4^3$ sub-boxes. We plot the PDF of the logarithmic mean of the NN-PDF, $l_1$ in Fig.~\ref{fig:error}. We choose the sub-samples to have  100 Mpc/h lengths for both the MDPL and the SMDPL. In this figure, we plot the normal fitting distributions with dashed yellow for MDPL. The dot-dashed green line is for the SMDPL. The plot ensures the errors are Gaussian. Also, results from the MDPL and the SMDPL agree with each other.\\
{{To further study the effect of boundary condition, we consider only halos inside a smaller box in the original simulation with the same centre. The length of the new box is $L-2l_b$ where $L$ is the original box size. However, the neighbours of halos which are inside the smaller box can still be in the original box. This implies that the boundary condition of the smaller box is more realistic than the original one (as we mentioned in Section~(\ref{Sec3: details}) we do not consider a periodic boundary condition in this study)}} \\
\begin{table*}
\centering
\caption{{{The moments of the SC-PDF for the case $M_{\lim}>10^{12}M_{\odot}$ and redshift $z=0$, the SMDPL simulation, calculated by equation~(\ref{Eq: moments}). The  quantity $l_b$ is shown in the first column. In columns 2, 3, 4, 5 the quantities $s_1$, $s_2$, $\tilde{s}_3$ and $\tilde{s}_4$ are respectively represented.} }}
\label{Ta: boundary-SC}
\begin{tabular}{ccccc}
\hline
$l_b$(Mpc/h) & $s_1$ & $s_2$ & $\tilde{s}_3$ & $\tilde{s}_4$ \\
\hline
\hline
0  & $1.171 \pm 0.032$ & $0.563 \pm 0.022$ & $0.713 \pm 0.082$ & $3.520 \pm 0.304$ \\ 
5  & $1.163 \pm 0.031$ & $0.554 \pm 0.021$ & $0.692 \pm 0.084$ & $3.461 \pm 0.312$ \\
15 & $1.159 \pm 0.030$ & $0.553 \pm 0.020$ & $0.690 \pm 0.077$ & $3.451 \pm 0.294$ \\
30 & $1.156 \pm 0.036$ & $0.548 \pm 0.023$ & $0.661 \pm 0.077$ & $3.341 \pm 0.246$ \\ 
\hline
\end{tabular}
\end{table*}
\begin{table*}
\centering
\caption{{{The logarithmic moments of the NN-PDF for the case $M_{\lim}>10^{12}M_{\odot}$ and redshift $z=0$, the SMDPL simulation, calculated by equation~(\ref{Eq: logmoments}). The quantity $l_b$ is shown in column 1. In columns 2, 3, 4, 5  the quantities $l_1$, $l_2$, $\tilde{l}_3$ and $\tilde{l}_4$ are respectively represented.} }}
\label{Ta: boundary-NN}
\begin{tabular}{ccccc}
\hline
$l_b$(Mpc/h) & $l_1$ & $l_2$ & $\tilde{l}_3$ & $\tilde{l}_4$ \\
\hline
\hline
0  & $-0.649 \pm 0.010$ & $0.538 \pm 0.007$ & $0.076 \pm 0.027$ & $2.637 \pm 0.035$ \\ 
5  & $-0.654 \pm 0.010$ & $0.535 \pm 0.007$ & $0.068 \pm 0.026$ & $2.632 \pm 0.036$ \\ 
15 & $-0.655 \pm 0.012$ & $0.535 \pm 0.008$ & $0.068 \pm 0.030$ & $2.631 \pm 0.040$ \\ 
30 & $-0.655 \pm 0.017$ & $0.535 \pm 0.008$ & $0.070 \pm 0.033$ & $2.633 \pm 0.044$ \\ 
\hline
\end{tabular}
\end{table*}
{{In Tables~(\ref{Ta: boundary-SC},\ref{Ta: boundary-NN}), we represent moments of the SC-PDF and log-moments of the NN-PDF, for different $l_b$, respectively. For this sample $r_{\star}=3.8$ Mpc/h. Accordingly, we conclude the effect of boundary condition is negligible for the case which $l_b=30$ Mpc/h. In Tables (\ref{Ta: boundary-SC},\ref{Ta: boundary-NN}), the moments and log-moments agree with each-other and difference between the moments is always smaller than our conservative errors. For the moments of SC-PDF, difference is a few percent, while for the log-moments of the NN-PDF, this difference is below one percent.\\
We conclude that the effects induced from boundary condition, is negligible in our work, when we consider conservative errors. However, for a more accurate analysis, one should consider more realistic boundary conditions in simulations, specially for the SC-PDF. Furthermore, we find that, the SC-PDF does not depend on the box size, seed number, and more importantly, the same linear relation between $s_1$ and $s_2$ exist for the other simulations (MDPL and MDPL2). In large-box simulations and small mass scales, the number of halos is large. And when the statistical errors are comparable with systematic errors,  more accurate methods are necessary.}}
\section{General behaviour of the SC-PDF and the NN-PDF} \label{App_Gen}
As we discussed in the main text, we focus on the normal and log-normal distributions as indicators of the linear and non-linear scales.  We calculated the moments of the SC-PDF in Table~(\ref{Ta: SCF}) and the log-moments of the NN-PDF in Table~(\ref{Ta: NND}). We showed that the SC-PDF and the NN-PDF follow approximately the normal and log-normal distribution, respectively. However, we can ask the question vice versa. What about the log-moments of the SC-PDF and moments of the NN-PDF?
Note that the kurtosis of the normal distribution is three. We use this property to distinguish the normal and log-normal distributions. 
In this appendix, we show that the NN-PDF, due to its large kurtosis, is far from a normal distribution. For the SC-PDF the kurtosis of the normal and log-normal distribution is almost the same regarding the confidence level. 
\begin{table*}
\centering
\caption{The logarithmic moments of the SC-PDF related to samples of Table~(\ref{Ta: DM samples}) calculated by equation~(\ref{Eq: logmoments}). The condition on mass and redshift is presented in columns one and two, respectively. $l_1$, $l_2$, $\tilde{l}_3$ and $\tilde{l}_4$ are reported in following columns. }
\label{Ta: SCF1}
\begin{tabular}{cccccc}
\hline
$\log(M_{\lim}/M_{\odot}h)$ & redshift(z) & $l_1$ & $l_2$ & $\tilde{l}_3$ & $\tilde{l}_4$ \\
\hline
\hline
11 & 0 & $0.091 \pm 0.019$ & $0.569 \pm 0.004$ & $-0.703 \pm 0.02$ & $3.907 \pm 0.059$ \\
11 & 0.5 & $0.082 \pm 0.018$ & $0.565 \pm 0.003$ & $-0.717 \pm 0.021$ & $3.957 \pm 0.068$ \\
11 & 1 & $0.079 \pm 0.018$ & $0.561 \pm 0.004$ & $-0.729 \pm 0.021$ & $3.958 \pm 0.065$ \\
\hline
11.5 & 0 & $0.059 \pm 0.021$ & $0.554 \pm 0.005$ & $-0.735 \pm 0.025$ & $3.979 \pm 0.085$ \\
11.5 & 0.5 & $0.053 \pm 0.022$ & $0.549 \pm 0.005$ & $-0.745 \pm 0.03$ & $4.009 \pm 0.122$ \\
11.5 & 1 & $0.054 \pm 0.022$ & $0.546 \pm 0.003$ & $-0.756 \pm 0.032$ & $4.015 \pm 0.103$ \\
\hline
12 & 0 & $0.029 \pm 0.026$ & $0.541 \pm 0.007$ & $-0.768 \pm 0.043$ & $4.04 \pm 0.154$ \\
12 & 0.5 & $0.028 \pm 0.025$ & $0.536 \pm 0.006$ & $-0.781 \pm 0.052$ & $4.089 \pm 0.238$ \\
12 & 1 & $0.029 \pm 0.025$ & $0.531 \pm 0.006$ & $-0.81 \pm 0.049$ & $4.187 \pm 0.242$ \\
\hline
12.5 & 0 & $0.004 \pm 0.03$ & $0.528 \pm 0.011$ & $-0.815 \pm 0.095$ & $4.284 \pm 0.475$ \\
12.5 & 0.5 & $0.004 \pm 0.028$ & $0.523 \pm 0.011$ & $-0.825 \pm 0.092$ & $4.26 \pm 0.522$ \\
12.5 & 1 & $0.006 \pm 0.035$ & $0.519 \pm 0.013$ & $-0.849 \pm 0.113$ & $4.267 \pm 0.415$ \\
\hline
13 & 0 & $-0.012 \pm 0.042$ & $0.513 \pm 0.017$ & $-0.82 \pm 0.161$ & $4.228 \pm 0.613$ \\
13 & 0.5 & $-0.014 \pm 0.038$ & $0.504 \pm 0.021$ & $-0.831 \pm 0.173$ & $4.244 \pm 1.005$ \\
13 & 1 & $-0.005 \pm 0.048$ & $0.501 \pm 0.021$ & $-0.929 \pm 0.198$ & $4.613 \pm 1.063$ \\
\hline
13.5 & 0 & $-0.028 \pm 0.061$ & $0.498 \pm 0.034$ & $-0.873 \pm 0.247$ & $4.518 \pm 1.255$ \\
13.5 & 0.5 & $-0.043 \pm 0.052$ & $0.489 \pm 0.036$ & $-0.847 \pm 0.159$ & $3.983 \pm 0.79$ \\
13.5 & 1 & $-0.05 \pm 0.073$ & $0.47 \pm 0.049$ & $-0.778 \pm 0.323$ & $3.947 \pm 1.462$ \\
\hline
14 & 0 & $-0.08 \pm 0.074$ & $0.48 \pm 0.055$ & $-0.875 \pm 0.432$ & $4.157 \pm 1.542$ \\
14 & 0.5 & $-0.071 \pm 0.096$ & $0.471 \pm 0.096$ & $-0.857 \pm 0.588$ & $3.918 \pm 2.151$ \\
14 & 1 & $-0.133 \pm 0.145$ & $0.41 \pm 0.149$ & $-0.476 \pm 0.648$ & $2.907 \pm 1.132$ \\
\hline
\end{tabular}
\end{table*}
\begin{table*}
\centering
\caption{The moments of the NN-PDF related to the samples of Table~(\ref{Ta: DM samples}) calculated by equation~(\ref{Eq: moments}). The condition on mass and redshift are presented in columns one and two respectively. $s_1$, $s_2$, $\tilde{s}_3$ and $\tilde{s}_4$ are reported in following columns.}
\label{Ta: NND1}
\begin{tabular}{cccccc}
\hline
$\log(M_{\lim}/M_{\odot}h)$ & redshift(z) & $s_1$ & $s_2$ & $\tilde{s}_3$ & $\tilde{s}_4$ \\
\hline
\hline
11 & 0 & $0.566 \pm 0.007$ & $0.347 \pm 0.007$ & $1.82 \pm 0.035$ & $8.253 \pm 0.301$ \\
11 & 0.5 & $0.567 \pm 0.006$ & $0.345 \pm 0.007$ & $1.832 \pm 0.04$ & $8.228 \pm 0.347$ \\
11 & 1 & $0.563 \pm 0.006$ & $0.345 \pm 0.007$ & $1.862 \pm 0.034$ & $8.28 \pm 0.264$ \\
\hline
11.5 & 0 & $0.586 \pm 0.007$ & $0.347 \pm 0.008$ & $1.703 \pm 0.046$ & $7.458 \pm 0.326$ \\
11.5 & 0.5 & $0.585 \pm 0.007$ & $0.346 \pm 0.008$ & $1.741 \pm 0.055$ & $7.608 \pm 0.457$ \\
11.5 & 1 & $0.579 \pm 0.007$ & $0.348 \pm 0.008$ & $1.754 \pm 0.049$ & $7.468 \pm 0.341$ \\
\hline
12 & 0 & $0.604 \pm 0.008$ & $0.347 \pm 0.009$ & $1.589 \pm 0.082$ & $6.831 \pm 0.612$ \\
12 & 0.5 & $0.601 \pm 0.008$ & $0.348 \pm 0.01$ & $1.633 \pm 0.08$ & $6.892 \pm 0.673$ \\
12 & 1 & $0.592 \pm 0.01$ & $0.353 \pm 0.011$ & $1.642 \pm 0.061$ & $6.71 \pm 0.404$ \\
\hline
12.5 & 0 & $0.621 \pm 0.012$ & $0.35 \pm 0.015$ & $1.511 \pm 0.096$ & $6.254 \pm 0.708$ \\
12.5 & 0.5 & $0.617 \pm 0.012$ & $0.357 \pm 0.014$ & $1.555 \pm 0.11$ & $6.305 \pm 0.701$ \\
12.5 & 1 & $0.604 \pm 0.015$ & $0.366 \pm 0.015$ & $1.544 \pm 0.099$ & $6.121 \pm 0.647$ \\
\hline
13 & 0 & $0.639 \pm 0.021$ & $0.36 \pm 0.022$ & $1.442 \pm 0.135$ & $5.681 \pm 0.794$ \\
13 & 0.5 & $0.638 \pm 0.023$ & $0.375 \pm 0.022$ & $1.459 \pm 0.15$ & $5.705 \pm 0.995$ \\
13 & 1 & $0.625 \pm 0.03$ & $0.39 \pm 0.03$ & $1.341 \pm 0.234$ & $5.086 \pm 1.609$ \\
\hline
13.5 & 0 & $0.662 \pm 0.035$ & $0.378 \pm 0.032$ & $1.319 \pm 0.271$ & $5.095 \pm 1.536$ \\
13.5 & 0.5 & $0.655 \pm 0.049$ & $0.385 \pm 0.035$ & $1.171 \pm 0.293$ & $4.423 \pm 1.615$ \\
13.5 & 1 & $0.664 \pm 0.057$ & $0.425 \pm 0.046$ & $1.088 \pm 0.341$ & $3.977 \pm 1.227$ \\
\hline
14 & 0 & $0.705 \pm 0.053$ & $0.399 \pm 0.047$ & $1.119 \pm 0.424$ & $4.356 \pm 1.94$ \\
14 & 0.5 & $0.718 \pm 0.082$ & $0.423 \pm 0.055$ & $0.854 \pm 0.381$ & $3.341 \pm 1.117$ \\
14 & 1 & $0.772 \pm 0.165$ & $0.435 \pm 0.105$ & $0.363 \pm 0.559$ & $2.138 \pm 0.833$ \\
\hline
\end{tabular}
\end{table*}
In Tables~(\ref{Ta: SCF1}),(\ref{Ta: NND1}), we calculate the logarithmic moments of the SC-PDF and the moments of the NN-PDF. Noting column four of Table~(\ref{Ta: SCF1}), the log-kurtosis $\tilde{l}_4$ is larger than 3.9 for all the cases except the large mass case. We can compare this number with column four of Table~(\ref{Ta: SCF}), in which the kurtosis of the SC-PDF, $\tilde{s}_4$, is smaller than 3.9 for all cases. However, considering errors, for most cases, two numbers overlap. So there is an opportunity to approximate the SC-PDF with a logarithmic distribution. The situation is more clear for NN-PDF. Comparing column four between Table~(\ref{Ta: NND1}) and Table~(\ref{Ta: NND}), log-kurtosis is around 2.6 but range of kurtosis is from 8.2 to 2.1. So NN-PDF is nearly log-normal can't be approximated with a normal distribution.\\
\begin{figure*}
\includegraphics[ width=1\textwidth]{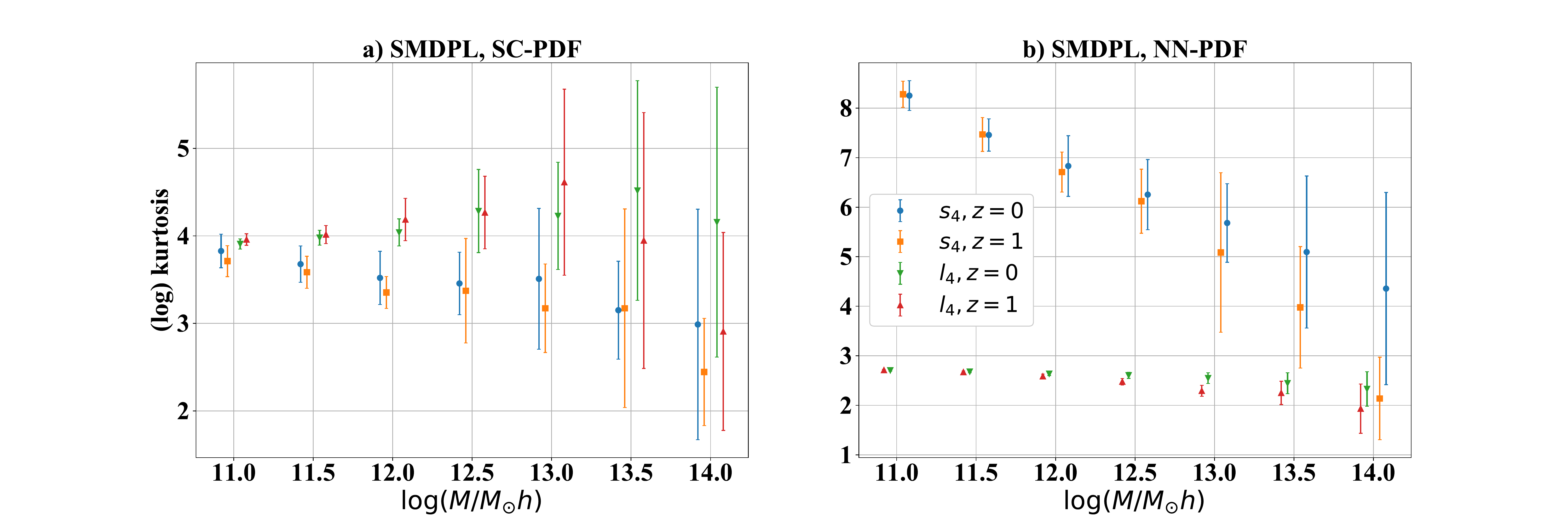}
\caption{Kurtosis and logarithmic kurtosis of both (a) the SC-PDF and (b) the NN-PDF are plotted for SMDPL simulation versus mass scale. $\tilde{s}_4$ is plotted with blue circles for $z=0$ and orange squares show $z=1$. Also $\tilde{l}_4$ is plotted with green down-pointing triangle for $z=0$ and red up-pointing triangle for $z=1$. Mass scales are shifted slightly for clarity.}
\label{fig:l4}
\end{figure*}
In Fig.~\ref{fig:l4}, we plot the kurtosis and log-kurtosis corresponding to halos from the SMDPL. We present the SC-PDF in panel (a) and the NN-PDF in panel (b). In  the figure, the blue circles show $z=0$ ($\tilde{s}_4$), orange squares show $z=1$ ($\tilde{s}_4$), green down-pointing triangle show $z=0$ ($\tilde{l}_4$) and red up-pointing triangle show $z=1$ ($\tilde{l}_4$). (again, similar to Fig.~\ref{fig:s}, we shifted mass scales slightly for clarity). The situation for the NN-PDF is certain when the logarithmic kurtosis deviates from the normal kurtosis completely. For the SC-PDF for  scales ($10^{11}$ and $10^{12}$), $\tilde{s}_4$ is lower than $\tilde{l}_4$ and it is closer to three comparing to $\tilde{l}_4$. For the other mass limits, considering the error bars, we can not distinguish the two classes. However, considering mean kurtosis, for all cases, $\tilde{s}_4$ is lower than $\tilde{l}_4$. Accordingly, in the main text, we report that the SC-PDF is a nearly normal distribution.


\bsp	
\label{lastpage}
\end{document}